\documentclass[twocolumn,showpacs,preprintnumbers,amsmath,amssymb,pra]{revtex4}

\usepackage{graphicx}
\usepackage{dcolumn}
\usepackage{bm}
\usepackage{epsfig}

\newcommand{\be}{\begin{eqnarray}}
\newcommand{\ee}{\end{eqnarray}}

\begin{document}

\title{Analytical thermodynamics of a strongly attractive
three-component Fermi gas in one dimension}
\author{Peng He$^{1,2}$, Xiangguo Yin$^{1}$, Xiwen Guan$^{2}$, Murray
  T. Batchelor$^{2,3}$ and Yupeng Wang$^{1}$}
\affiliation{${1}$ Beijing
  National Laboratory for Condensed Matter Physics, Institute of
  Physics, Chinese Academy of Sciences, Beijing 100190, P. R. China}
  \affiliation{${2}$ Department of Theoretical Physics, Research School of
  Physics and Engineering, Australian National University,
  Canberra ACT 0200, Australia}
\affiliation{${3}$ Mathematical Sciences Institute, Australian
  National University, Canberra ACT 0200, Australia}

\begin{abstract}
Ultracold three-component atomic Fermi gases in one dimension are
expected to exhibit rich physics due to the presence of trions and
different pairing states.
Quantum phase transitions from the trion state into a paired phase
and a normal Fermi liquid occur at zero temperature.
We derive the analytical thermodynamics of strongly attractive
three-component one-dimensional fermions with $SU(3)$ symmetry via
the thermodynamic Bethe ansatz method in unequal Zeeman splitting
fields $H_1$ and $H_2$.
We find explicitly that for low temperature the system acts like
either a two-component or a three-component Tomonaga-Luttinger
liquid dependent on the system parameters.
The phase diagrams for the chemical potential and specific heat are
presented for illustrative values of the Zeeman splitting.
We also demonstrate that crossover between different
Tomonaga-Luttinger liquid phases exhibit singular behaviour in
specific heat and entropy as the temperature tends to zero.
Beyond Tomonaga-Luttinger liquid physics, we obtain the equation of
state which provides a precise description of universal
thermodynamics and quantum criticality  in  three-component strongly
attractive  Fermi gases.
\end{abstract}

 \maketitle

\section{Introduction}

The ongoing experimental advances in realizing degenerate quantum
gases in low dimensions
\cite{TGexp,Toshiya,Moritz05,Druten,Haller,Hulet} offer a new and
compelling motivation for the further study of quantum many-body
systems via exact schemes such as the Bethe Ansatz (BA) and low
energy effective field theory \cite{Giamarchi-b}.
Reducing the dimensionality in a quantum system can have striking
consequences.
The one-dimensional (1D) many-body systems
\cite{Giamarchi-b,Takahashi} possess unique many-body correlation
effects which are different from their higher dimensional
counterparts.
These include the phenomena of spin-charge separation, universal
thermodynamics and quantum criticality.

A recent scheme for mapping out physical properties of homogeneous
systems by using the inhomogeneity of the trap \cite{Ho-Zhou} has
been successfully applied to experimental measurements on the
thermodynamics of interacting fermions with a wide range of tunable
interactions \cite{Salomon,Horikoshi}.
Moreover, further experimental advances with ultracold atoms allow
the exploration of three-component Fermi gases in the entire
parameter space of trions, dimers and free atoms
\cite{Li,efimov-trimer,Li2}.
This provides a promising opportunity to experimentally explore
universal thermodynamics and quantum critical behaviour of strongly
interacting Fermi gases with high spin symmetries in 1D.
In this context, the thermodynamics of 1D attractively interacting
fermions \cite{Takahashi} has been receiving growing interest
\cite{Guan2007prb,Mueller,Bolech,Erhai}.

For spin-1/2 fermions with attractive interaction there are three
quantum phases at zero temperature: the fully paired phase which is
a quasi-condensate with zero polarization $p$, the fully polarized
(normal) phase with $p = 1$, and the partially polarized (1D FFLO)
phase where $0 < p < 1 $ at zero temperature
\cite{Orso,Hu,Guan2007prb}.
This theoretical prediction of the phase diagram for 1D fermions was
recently confirmed experimentally by R. Hulet's group at Rice
University \cite{Hulet}.
In addition, it was recently proved \cite{Erhai} that at low
temperatures, the physics of the gapless phase belongs to the
universality class of a two-component Tomonaga-Luttinger liquid
(TLL).
However, from the theoretical point of view, understanding the
thermodynamics of multi-component Fermi gases with higher spin
symmetry imposes a number  of challenges
\cite{Ho,Tsvelik,Zhang,Wang,Guan}.

For multi-component interacting Fermi gases, the phase diagrams
become more complicated in the presence of magnetic fields due to
the richer number of quantum phases.
In contrast to the two-component Fermi gas,
\cite{Orso,Hu,Guan2007prb} three-component ultracold fermions give
rise to quantum phase transitions from a three-body bound state of
``trions" into the BCS pairing state and a normal Fermi liquid
\cite{Rapp,Lecheminant,Demler,Guan2008prl,Thai,Silva,Errea,AnnPhys,Suga2009arXiv,Angela}.
The zero temperature phase diagrams of the BA integrable 1D
three-component Fermi gas with $SU(3)$ symmetry have been worked out
from the dressed energy equations \cite{Guan2008prl,Angela}.
It was found that Zeeman splittings can drive transitions between
exotic phases of trions, bound pairs, a normal Fermi liquid and
mixtures of these phases, see Fig.~\ref{fig:phases}.
It is thus very worthwhile to map out such zero temperature phase
diagrams to the inhomogeneity of the trap at finite temperatures.

In this paper, we investigate the finite temperature thermodynamic
properties of 1D three-component fermions with unequal Zeeman
splitting by means of the exact thermodynamic Bethe ansatz (TBA)
solution.
We prove that at low temperatures the system behaves like either  a
two-component or a three-component TLL in certain regimes.
Exact finite temperature phase diagrams are demonstrated for
illustrative values of the Zeeman splitting parameters.
Quantum criticality with respect to the specific heat and entropy as
the temperature tends to zero is discussed.
The equation of state obtained provides an exact description of the
thermodynamics and quantum critical behaviour of three-component
composite fermions which can possibly be tested in experiments with
ultracold atoms.

This paper is set out as follows. In Section II, we present the
model and the exact BA solution. We also derive the TBA equations
for the thermodynamics.  In Section III, we derive the low
temperature thermodynamics by the Sommerfeld expansion method. The
universal multi-component TLL phases are identified. In Section IV,
we present the equation of state in terms of polylogarithm functions
from which the quantum phase diagrams can be mapped out.
Concluding remarks are given in Section V.
Detailed working is given in the appendices.
The derivation of the TBA equations is presented in detail in
Appendix A.
In Appendices B and C, the iteration method is used to derive
relevant results for the TBA and the thermodynamics.

\section{The Model and the thermodynamic Bethe ansatz solution}
\label{sec:Ham}

We consider a 1D system of $N$ fermions of mass $m$ with spin
independent $\delta $-function potential interaction and are
constrained to a line of length $L$ with periodic boundary
conditions.
The fermions can occupy three possible hyperfine levels ($\left\vert
1\right\rangle $, $\left\vert 2\right\rangle $ and $\left\vert
3\right\rangle $) with particle number $N^{1}$, $N^{2}$ and $N^{3}$,
respectively.
The system can be described by the Hamiltonian
\cite{Sutherland,Takahashi-2}
\begin{equation}
\mathcal{H}_0=-\frac{\hbar ^{2}}{2m}\sum_{i=1}^{N}\frac{\partial
^{2}}{
\partial x_{i}^{2}}+g_{1D}\sum_{1\leq i<j\leq N}\delta
(x_{i}-x_{j})+E_Z \label{Ham}
\end{equation}
where we have included the Zeeman energy
$E_Z=\sum_{i=1}^{3}N^{i}\epsilon^{i}_Z(\mu_B^{i},B)$.
The spin-independent contact interaction $g_{\rm 1D}$ applies
between fermions with different hyperfine states so that the number
of fermions in each spin state is conserved.  The inter-component
interaction $g_{1D}$ is positive for repulsive interaction and
negative for attractive interaction.
For simplicity, we define the interaction strengths as
$c=mg_{1D}/\hbar ^{2}$ and the dimensionless parameter $\gamma
=c/n$, where $n=N/L$ is the linear density, and set $\hbar =2m=1$.
Although these conditions appear rather restrictive, it is possible
to tune scattering lengths between atoms in different low sublevels
to form nearly $SU(3)$ degeneracy Fermi gases via broad Feshbach
resonances \cite{Li,efimov-trimer,Li2}.

In the above equation, the Zeeman energy levels $\epsilon^{i}_Z $
are determined by the magnetic moments $\mu _{B}^{i}$ and the
magnetic field $B$. By convention, particle numbers in each of the
hyperfine states satisfy the relation $N^{1}\geq N^{2}\geq N^{3}$.
Thus the particle numbers of unpaired fermions, pairs, and trions
are respectively given by $N_{1}=N^{1}-N^{2}$, $N_{2}=N^{2}-N^{3}$
and $N_{3}=N^{3}$ for the attractive regime.

In order to simplify calculations in the study of population
imbalance, we rewrite the Zeeman energy as
$E_Z=-H_{1}N_{1}-H_{2}N_{2}+N\bar{\epsilon}$ where the unequally
spaced Zeeman splitting in three hyperfine levels can be
characterized by two independent parameters $H_1 = \bar{\epsilon} -
\epsilon^{1}_Z(\mu_B^{1},B)$ and $H_2 = \epsilon^{3}_Z(\mu_B^{3},B)-
\bar{\epsilon}$, with
$\bar{\epsilon}=\sum_{i=1}^3\epsilon^{i}_Z(\mu_B^{i},B)/3$ the
average Zeeman energy.
Pure Zeeman splitting (equally-spaced splitting), i.e.
$H_{1}=H_{2}$, leads to a smooth phase transition from trions into
the normal Fermi liquid.
On the other hand, unequally-spaced Zeeman splitting can lead to
quantum phase transitions from trions to the fully-paired phase and
to a mixture of pairs and single atoms, see Fig.~\ref{fig:phases}.

The Hamiltonian (\ref{Ham}) exhibits a symmetry of $U(1) \times
SU(3)$, where $U(1)$ and $ SU(3)$ describe the charge and spin
degrees of freedom.
This model was solved long ago by means of the nested Bethe ansatz
\cite{Sutherland,Takahashi-2}.
The energy eigenspectrum is given in terms of the quasimomenta
$\left\{k_j\right\}$ of the $N$ fermions by
\begin{equation}
E= \sum_{j=1}^N k_j^2
\end{equation}
which satisfy the BA equations
\cite{Sutherland,Takahashi-2}
\begin{eqnarray}
e^{\mathrm{i} k_{j}L}&=&\prod_{\ell=1}^{M_{1}}\frac{k_{j}-\Lambda
_{\ell}+\mathrm{i}c/2}{k_{j}-\Lambda _{\ell}-\mathrm{i}c/2},
\nonumber\\
\prod_{j=1}^{N}\frac{\Lambda _{\ell}-k_{j}+\mathrm{i}c/2}{\Lambda
_{\ell}-k_{j}-\mathrm{i}c/2}&=&-\prod_{\alpha
=1}^{M_{1}}\frac{\Lambda _{\ell}-\Lambda _{\alpha
}+\mathrm{i}c}{\Lambda _{\ell}-\Lambda _{\alpha
}-\mathrm{i}c}\nonumber\\
&& \times \prod_{m=1}^{M_{2}}\frac{\Lambda _{\ell}-\lambda
_{m}-\mathrm{i}c/2}{\Lambda _{\ell}-\lambda
_{m}+\mathrm{i}c/2}, \nonumber\\
 \prod_{\ell=1}^{M_{1}}\frac{\lambda
_{m}-\Lambda _{\ell}+\mathrm{i}c/2}{\lambda _{m}-\Lambda
_{\ell}-\mathrm{i}c/2}&=&-\prod_{\beta =1}^{M_{2}}\frac{\lambda
_{m}-\lambda _{\beta }+\mathrm{i}c}{\lambda _{m}-\lambda _{\beta
}-\mathrm{i}c}.  \label{BAE}
\end{eqnarray}
Here $j=1,\ldots ,N$, $\ell =1,\ldots ,M_{1}$, $m=1,\ldots ,M_{2}$
with quantum numbers $M_{1}=N^{2}+N^{3}$ and $M_{2}=N^{3}$.
The parameters $\left\{ \Lambda _{\ell},\lambda _{m}\right\}$ are
the rapidities for the internal hyperfine spin degree of freedom.

In the thermodynamic limit, $N,L\rightarrow\infty$ with $n$ finite,
the sets of solutions $\{k_{j}\}$, $\{\Lambda_{\ell}\}$ and
$\{\lambda_{m}\}$ of the BA equations (\ref{BAE}) are of certain
forms, as discussed in Appendix A.
For attractive interaction the quasimomenta $\{k_{j}\}$ can form
two-body and three-body charge bound states, which give a natural
description of composite fermions,  and can also be real
\cite{Takahashi-2,Guan2008prl}.
However, the rapidities $\{\Lambda_{\ell}\}$ and $\{\lambda_{m}\}$
can form complex spin-strings characterizing the spin wave
fluctuations at finite temperatures.

In the thermodynamic limit, the grand partition function
\cite{Yang1969,Takahashi}
$Z=\mathrm{tr}(\mathrm{e}^{-H/T})=\mathrm{e}^{-G/T}$ is given in
terms of the Gibbs free energy
\begin{eqnarray}
G &=& E -\mu N + E_{\rm Z} -TS \nonumber\\
&=& E-\mu N-H_{1}N_{1}-H_{2}N_{2}-TS, \label{Gibbs}
\end{eqnarray}
where the chemical potential $\mu$, the Zeeman energy $E_{\rm Z}$
and the entropy $S$ are given in terms of the densities of unpaired
fermions, charge bound states, trions and spin-strings,  which are
all subject to the BA equations (\ref{BAE}).
The equilibrium states are determined by minimizing the Gibbs free
energy, which gives rise to a set of coupled nonlinear integral
equations -- the TBA equations for the dressed energies
$\varepsilon _{a} (a=1,2,3)$, which are derived for this model in
Appendix A, with final result
\begin{eqnarray}
\varepsilon _{1} (k)  &=&k^{2}-\mu -H_{1} + \, Ta_{1}\ast \ln
(1+e^{-{\varepsilon _{2}}/{T}})(k)   \nonumber \\
&&+ \, T a_{2}\ast \ln (1+e^{-\varepsilon _{3}/T})(k) \nonumber \\
&&- \, T\sum_{n=1}^\infty a_{n}\ast \ln (1+\xi _{n}^{-1})(k), \nonumber  \\
\varepsilon _{2}(k) &=& 2k^{2}-\frac12{c^{2}} -2\mu -H_{2} \nonumber\\
&&+ \, Ta_{1}\ast \ln (1+e^{-{\varepsilon _{1}}/{T}})(k) \nonumber\\
&&+ \, Ta_{2}\ast \ln (1+e^{-{\varepsilon _{2}}/T}) (k) \nonumber  \\
&&+ \, T(a_{1}+a_{3})\ast \ln (1+e^{-{\varepsilon _{3}}/T})(k) \nonumber   \\
&&- \, T\sum_{n=1}^\infty a_{n}\ast \ln (1+\zeta _{n}^{-1}) (k), \nonumber \\
 \varepsilon _{3}(k)  &=&3 k^{2}-2{c^{2}}-3\mu
+Ta_{2}\ast \ln (1+e^{-\varepsilon _{1}/T}) (k) \nonumber   \\
&&+ \, T(a_{1}+a_{3})\ast \ln (1+e^{-{\varepsilon _{2}}/{T}}) (k)  \nonumber   \\
&&+ \, T(a_{2}+a_{4})\ast \ln (1+e^{-{\varepsilon_{3}}/{T}}) (k).
 \label{TBA}
\end{eqnarray}
Here the quantity
\begin{equation}
a_{m}\left( x\right) =\frac{1}{2\pi }\frac{m\left\vert c\right\vert
}{\left( mc/2\right) ^{2}+x^{2}}
\end{equation}
and $\ast $ denotes the convolution,
\begin{equation}
(a\ast b)(x)=\int a(x-y)b(y)dy.
\end{equation}

The spin string parameters $\xi_{n}:= \sigma _{n}^{h}/\sigma _{n}$
and $\zeta _{n}:= \tau _{n}^{h}/\tau _{n}$ associated with particle
and hole densities of string length $n$ in $\Lambda$  and $\lambda$
parameter spaces satisfy the  string TBA equations
\begin{eqnarray}
\ln \xi _{n}\left( \Lambda \right)
&=&\frac{n(2H_{1}-H_{2})}{T}+a_{n}\ast \ln (1+e^{-{\varepsilon
_{1}}/{T}})\left( \Lambda \right) \nonumber \\
&&+\sum_{m}T_{mn}\ast
\ln (1+\xi _{m}^{-1})\left( \Lambda \right) \nonumber \\
&&-\sum_{m}S_{mn}\ast \ln (1+\zeta _{m}^{-1})\left( \Lambda
\right),\nonumber \\
\ln \zeta _{n}\left( \lambda \right)
&=&\frac{n(2H_{2}-H_{1})}{T}+a_{n}\ast \ln (1+e^{-{\varepsilon
_{2}}/{T}})\left( \lambda \right) \nonumber \\
&&+\sum_{m} T_{mn}\ast \ln
(1+\zeta _{m}^{-1})\left( \lambda \right) \nonumber \\
&&-\sum_{m}S_{mn}\ast \ln (1+\xi _{m}^{-1})\left( \lambda \right).
\label{TBA-string}
\end{eqnarray}
The functions $T_{mn}$ and $ S_{mn}$ are as defined in Appendix A.

In the thermodynamic limit, the pressure $p$ is defined in terms of
the Gibbs energy (\ref{Gibbs}) by $p\equiv -(\partial G/\partial
L)$, which includes three parts, $p^{(1)}$, $p^{(2)}$ and $p^{(3)}$,
for the pressure of unpaired fermions, pairs and trions,
respectively,
where
\begin{equation}
p^{(a)}=\frac{aT}{2\pi }\int dk\ln \left(1+e^{-\varepsilon_{a}\left(
k\right)/{T}}\right).\label{pressure}
\end{equation}
Here we have set the Boltzmann constant $k_{B}=1$.

 The TBA equations
 (\ref{TBA}) are expressed in terms of the dressed energies
 $\varepsilon_{1}(k)$, $\varepsilon_{2}(k)$ and $\varepsilon_{3}(k) $
 for unpaired fermions, pairs and trions, respectively. The dressed
 energies  are seen to depend not only on the chemical potential $\mu$
 and the external fields $H_1$ and $H_2$ but also on the interactions
 among themselves as well as the spin fluctuations characterized by the
 spin-strings (\ref{TBA-string}).
We clearly see that spin
 fluctuations are ferromagnetically coupled to the dressed energies
 for unpaired fermions and pairs.  There is no such spin fluctuation
 coupled to the dressed energy of the spin neutral trion states.  The
 TBA equations play the central role in the investigation of thermodynamic
 properties of exactly solvable models at finite temperature.  They also provides a convenient
 formalism to analyze quantum phase transitions and magnetic effects
 in the presence of external fields at zero temperature \cite{review}.

\section{Universal Tomonaga-Luttinger liquid phases}
\label{sec:TLL}

The TBA equations (\ref{TBA}) and (\ref{TBA-string}) involve an
infinite number of coupled nonlinear integral equations which
hinders access to the thermodynamics from both the analytical and
numerical points of view.
In the strong coupling regime, the dressed energies
$\varepsilon_{a}(k)$ with $a=1,\,2,\,3$ marginally depend on each
other. The spin string contributions to thermal fluctuations in the
strong coupling regime and at low temperatures, i.e. $T\ll H_1$ and
$T \ll H_2$,  are negligible.
In this temperature regime, the TBA equations (\ref{TBA}) can be
sorted as
\begin{eqnarray}
\varepsilon _{a}(k) &\approx &a\,k^{2}-A^{(a)},\qquad a=1,\,2,\,3,
\label{epsilon3}
\end{eqnarray}
in terms of the dressed chemical potentials
\begin{eqnarray}
A^{(1)} &=&\mu +H_{1}-\frac{2}{|c|}p^{(2)}-\frac{2}{3|c|}p^{(3)},
\nonumber \\
A^{(2)} &=&2\mu +\frac12 {c^{2}}
+H_{2}-\frac{4}{|c|}p^{(1)}-\frac{1}{|c|}p^{(2)}-
\frac{16}{9|c|}p^{(3)}, \nonumber\\
A^{(3)} &=&3\mu
+2c^{2}-\frac{2}{|c|}p^{(1)}-\frac{8}{3|c|}p^{(2)}-\frac{1}{|c|}
p^{(3)}. \label{AAA}
\end{eqnarray}
In this case we can directly calculate the pressure through
(\ref{pressure}), with result
\begin{equation}
p^{(a)}= \frac{\sqrt a}{\pi} \int_{0}^{\infty} \frac{
\sqrt{\varepsilon_a} d \varepsilon_a}{1+ \mathrm{e}^{(\varepsilon_a
-A^{(a)})/T}}
 \label{p_all}
\end{equation}
in terms of chemical potential $\mu $, temperature $T$ and external
fields $H_{1}$ and $H_{2}$.
Using  Sommerfeld expansion, we obtain  the pressure $p^{(a )}$ at
low temperatures,
\begin{equation}
p^{(a)} \approx \frac{2}{3}\sqrt{\frac{a}{{\pi ^{2}}}}\left( A^{(a
)}\right) ^{\frac{3}{2}}\left[ 1+\frac{\pi
^{2}}{8}\left(\frac{T}{A^{(a)}} \right)^{2}\right].  \label{p_gamma}
\end{equation}

The fields $H_1$ and $H_2$ may drive the system into a number of
different phases.
In order to extract the nature of the TLL physics from the low
temperature thermodynamics, we first consider the phase in which
trions, pairs and unpaired fermions coexist.
In this coexisting phase, we can apply Sommerfeld expansion under
the condition that the effective chemical potentials for trions,
pairs and unpaired fermions are greater than the temperature scale.
Iteration with the defining relations
\begin{eqnarray}
n =\frac{\partial p}{\partial \mu }, \,\,\, n_{1} =\frac{\partial
p}{\partial H_{1}}, \,\,\, n_{2} =\frac{\partial p}{\partial H_{2}},
\label{n}
\end{eqnarray}
leads to explicit forms for the pressure
\begin{widetext}
\begin{eqnarray}
p^{(1)} & \approx &\frac{2n_{1}^{3}\pi ^{2}}{3}\left(
1+\frac{12n_{2}}{|c|}+\frac{6n_{3}}{|c|}  +\frac{\pi
^{2}}{4}\left(\frac{T}{n_{1}^{2}{\pi^{2}}}\right)^{2}
\left[1-\frac{4n_2}{|c|}- \frac{2n_3}{|c|}\right]\right),  \\
p^{(2)} & \approx &\frac{n_{2}^{3}\pi ^{2}}{3}\left(
1+\frac{6n_{1}}{|c|}+\frac{3n_{2}}{|c|}+\frac{8n_{3}}{|c|}
+\frac{\pi ^{2}}{4}\left(
\frac{T}{\frac{n_{2}^{2}}{2}\pi ^{2}}\right)^{2}\left[1-\frac{2n_1}{|c|}-\frac{n_2}{|c|}- \frac{8n_3}{3|c|}\right]\right), \\
p^{(3)} &\approx&\frac{2n_{3}^{3}\pi
^{2}}{9}\left(1+\frac{2n_{1}}{|c|}+\frac{16n_2}{3|c|}+\frac{3n_3}{|c|}
+\frac{\pi ^{2}}{4}\left(
\frac{T}{\frac{n_{3}^{2}}{3}\pi^{2}}\right)^{2}\left[1-\frac{2n_1}{3|c|}-\frac{16n_2}{9|c|}-
\frac{n_3}{|c|}\right]\right). \label{p_se}
\end{eqnarray}
\end{widetext}
The detailed derivation is given in Appendix B.
For the total number of particles fixed, i.e.
 $n=n_{1}+2n_{2}+3n_{3}$, the free energy can be written as
\begin{eqnarray}
F &=&\mu n-p  \nonumber \\
&=&\mu ^{(1)}n_{1}+2\mu ^{(2)}n_{2}+3\mu ^{(3)}n_{3} \nonumber  \\
&&-H_{1}n_{1}-H_{2}n_{2}-\frac{c^2}{2}n_2-2c^{2}n_{3}-p,
\label{freeF}
\end{eqnarray}
where effective chemical potentials $\mu^{(a)}$s are given by
\begin{eqnarray}
\mu^{(1)}&=&\mu+H_1,\label{eff_mu1}\\
\mu^{(2)}&=&\mu+\frac14 {c^2}+\frac12{H_2},\label{eff_mu2}\\
\mu^{(3)}&=&\mu+\frac23{c^2}.\label{eff_mu3}
\end{eqnarray}

In order to see universal TLL physics, we calculate the leading low
temperature corrections to the free energy $F$.
Substituting $\mu ^{(a)}$ and $p^{(a)}$ into (\ref{freeF}), after
some lengthy calculation, we obtain the leading temperature
correction to the free energy
\begin{equation}
F \approx E_{0}-\frac{\pi T^{2}}{6}\left(
\frac{1}{v_{1}}+\frac{1}{v_{2}}+\frac{1}{v_{3}}\right),
\end{equation}
where the ground state energy is given by
\begin{equation}
E_0 = -H_1 n_1 - H_2 n_2 - \frac12 c^{2} n_2 - 2 c^{2} n_3
\end{equation}
and the velocities are
\begin{eqnarray}
v_{1} &\approx&2n_{1}\pi \left( 1+\frac{8}{|c|}n_{2}+\frac{4}{|c|}n_{3}\right),\nonumber  \\
v_{2} &\approx&4n_{2}\pi \left(
1+\frac{4}{|c|}n_{1}+\frac{2}{|c|}n_{2}+\frac{16}{3|c|}
n_{3}\right),\nonumber  \\
v_{3} &\approx&6n_{3}\pi \left(
1+\frac{4}{3|c|}n_{1}+\frac{32}{9|c|}n_{2}+\frac{2}{|c|}
n_{3}\right).\label{velocity}
\end{eqnarray}

The particle numbers $n_1$, $n_2$ and $n_3$ of different bound
states  can be obtained approximately by collecting terms up to
order $1/|c|$ in the expressions for the effective chemical
potentials $\mu^{(a)}$ in (\ref{eff_mu1})-(\ref{eff_mu3}) at zero
temperature,
\begin{widetext}
\begin{eqnarray}
\mu^{(1)} &\approx&n_{1}^{2}{\pi ^{2}}\left(1 +
\frac{2}{3|c|}\frac{n_{2}^{3}}{n_{1}^{2}} +
\frac{4}{27|c|}\frac{n_{3}^{3}}{n_{1}^{2}} + \frac{8}{|c|}n_{2} +
\frac{4}{|c|}n_{3} \right),\\
\mu^{(2)} & \approx& \frac{n_{2}^{2}}{4}\pi ^{2}\left(1+
\frac{16}{3|c|}\frac{n_{1}^{3}}{n_{2}^{2}} +
\frac{64}{81|c|}\frac{n_{3}^{3}}{n_{2}^{2}} + \frac{4}{|c|}n_{1}
+\frac{8}{3|c|}n_2 +\frac{16}{3|c|}n_{3} \right),  \\
\mu^{(3)} &\approx& \frac{n_{3}^{2}}{9}\pi^{2}\left(1+
\frac{4}{|c|}\frac{n_{1}^{3}}{n_{3}^{2}} +
\frac{8}{3|c|}\frac{n_{2}^{3}}{n_{3}^{2}}  + \frac{4}{3|c|}n_{1} +
\frac{32}{9|c|}n_{2} +\frac{8}{3|c|}n_3
 \right).
\end{eqnarray}
with final result
\begin{eqnarray}
n_1 &\approx& \frac{\sqrt{\mu^{(1)}}}{\pi} \left(1 -
\frac{8}{3\pi|c|}\frac{\left(\mu^{(2)}\right)^{\frac32}}{\mu^{(1)}}-
\frac{2}{\pi|c|} \frac{\left(\mu^{(3)}\right)^{\frac32}}{\mu^{(1)}}
- \frac{8}{\pi|c|}\sqrt{\mu^{(2)}} -
\frac{6}{\pi|c|}\sqrt{\mu^{(3)}} \right), \\
n_2 &\approx&  \frac{2\sqrt{\mu^{(2)}}}{\pi} \left(1 -
\frac{2}{3\pi|c|}\frac{\left(\mu^{(1)}\right)^{\frac32}}{\mu^{(2)}}-
\frac{8}{3\pi|c|} \frac{\left(\mu^{(3)}\right)^{\frac32}}{\mu^{(2)}}
- \frac{2}{\pi|c|}\sqrt{\mu^{(1)}}-
\frac{8}{3\pi|c|}\sqrt{\mu^{(2)}} -
\frac{8}{\pi|c|}\sqrt{\mu^{(3)}} \right), \\
n_3 &\approx& \frac{3\sqrt{\mu^{(3)}}}{\pi} \left(1 -
\frac{2}{9\pi|c|}
\frac{\left(\mu^{(1)}\right)^{\frac32}}{\mu^{(3)}}-
\frac{32}{27\pi|c|}
\frac{\left(\mu^{(2)}\right)^{\frac32}}{\mu^{(3)}} -
\frac{2}{3\pi|c|}\sqrt{\mu^{(1)}}-
\frac{32}{9\pi|c|}\sqrt{\mu^{(2)}} -
\frac{4}{\pi|c|}\sqrt{\mu^{(3)}} \right).
\end{eqnarray}
\end{widetext}

This result shows that strongly attractive three-component fermions
behave like a three-component TLL for the coexisting phase of
trions, pairs and unpaired fermions at low temperatures.
Similarly, we can extract the finite temperature corrections to the
free energy in other quantum phases.
For example, in the coexisting phase of trions and pairs, we have
the same universal form
\begin{equation}
F \approx E_{0}-\frac{\pi
T^{2}}{6}\left(\frac{1}{v_{2}}+\frac{1}{v_{3}}\right),
\end{equation}
where the velocities $v_2$ and $v_3$ have the same expressions as
that given in (\ref{velocity}) with $n_1=0$.
In the above equations, the free energy and the thermodynamics are
given in terms of the chemical potential and the effective Zeeman
fields $H_1$ and $H_2$. The chemical potential is convenient for
practical purposes in  experiments with cold atoms, where the
chemical potential is replaced by the harmonic potential
$\mu=\mu_0-\frac12 m \omega^2 x^2$. The relation between $\mu$ and
total particle number $n$ can be obtained from (\ref{n}).

Although there is no quantum phase transition in 1D many-body
systems at finite temperatures due to thermal fluctuations, we shall
show that the TLL phases persist for non-zero temperatures, as noted
in another context \cite{Maeda}.

\section{Thermodynamics at low temperatures}
\label{sec:EOS}

For strong attraction ($|\gamma| \gg 1$)  three-atom and two-atom
charge bound states
 can be  stable under  certain Zeeman fields.
The corresponding binding energies of the trions and pairs are given
 by $\varepsilon_{\rm t} = \hbar^2c^2/m$ and $\varepsilon_{\rm b} =
 \hbar^2c^2/4m$, respectively.  At high temperatures $T\sim
 \varepsilon_{\rm t }, \varepsilon_{\rm b}$, thermal fluctuations can
 break the charge bound states while spin fluctuations cannot be
 ignored.
 However, such spin fluctuations coupled to the channels of
 unpaired fermions and the spin-$1$ charge bound pairs are suppressed
 by large fields $H_1$ and $H_2$ at low temperatures.  In this regime,
 the spin string contributions to thermal fluctuations can be
 asymptotically calculated from the TBA equations (\ref{TBA}) and
 (\ref{TBA-string}), see Appendix C.
We have
\begin{eqnarray}
\varepsilon _{1}(k) & \approx &k^{2}-\mu
-H_{1}+\frac{2}{|c|}p^{(2)}+\frac{2}{3|c|} p^{(3)}  \nonumber \\
&&- \, T \,  \mathrm{e}^{-(2H_{1}-H_{2})/{T}} \mathrm{e}^{-{J_{1}}/{T}} I_{0}(\frac{J_{1}}{T}), \nonumber \\
\varepsilon _{2}(k) &\approx&2k^{2}-\frac{c^{2}}{2}-2\mu
-H_{2}+\frac{4}{|c|}p^{(1)}+
\frac{1}{|c|}p^{(2)} \nonumber  \\
&&+ \, \frac{16}{9|c|}p^{(3)} - \, T \,
\mathrm{e}^{-(2H_{2}-H_{1})/{T}}
\mathrm{e}^{-{J_{2}}/{T}} I_{0}(\frac{J_{2}}{T}),\nonumber  \\
\varepsilon _{3}(k) &\approx&3k^{2}-3\mu
-2c^{2}+\frac{2}{|c|}p^{(1)}+\frac{8}{3|c|} p^{(2)} \nonumber  \\
&& + \, \frac{1}{|c|}p^{(3)}, \label{DEC}
\end{eqnarray}
where $J_{1}={2}p_{1}/{|c|}$ and $J_{2}=p_{2}/{|c|}$ effective
spin-spin interactions and
\begin{equation}
I_{n}(z)=\frac{1}{\pi }\int_{0}^{\pi }e^{z\cos \theta }\cos
(n\theta)d\theta .
\end{equation}
We also see clearly that there is no such effective spin-spin
interaction for the spin-neutral trion bound state.

Using the formula (\ref{p_all}), we can write the pressure $p^{(a
)}$  in terms of the polylogarithm function, i.e.
\begin{equation}
p^{(a )}=-\sqrt{\frac{a }{4\pi }} \, T^{{3}/{2}} \,
\mathrm{{Li}}_{{3}/{2}}\left(-\mathrm{e}^{A^{(a )}/{T}}\right),
\label{ppp}
\end{equation}
for $a=1,\,2,\,3$, where the polylogarithm function is defined as
\begin{equation}
\mathrm{Li}_{1+s}(-\mathrm{e}^x)=-\frac{1}{\Gamma
(s+1)}\int_{0}^{\infty } \frac{k^{s}dk}{e^{k-x}+1}.
\end{equation}
To leading order, the functions $A^{(a)}$ are
\begin{eqnarray}
A^{(1)} &=&\mu +H_{1}-\frac{2}{|c|}p^{(2)}-\frac{2}{3|c|}p^{(3)}
\nonumber \\
&&+ \, T \,  \mathrm{e}^{-(2H_{1}-H_{2})/{T}} \mathrm{e}^{-{J_{1}}/{T}} I_{0}(\frac{J_{1}}{T}), \nonumber \\
A^{(2)} &=&2\mu
+\frac{c^{2}}{2}+H_{2}-\frac{4}{|c|}p^{(1)}-\frac{1}{|c|}p^{(2)}-
\frac{16}{9|c|}p^{(3)} \nonumber\\
&& + \, T \, \mathrm{e}^{-(2H_{2}-H_{1})/{T}}
\mathrm{e}^{-{J_{2}}/{T}} I_{0}(\frac{J_{2}}{T}),\nonumber  \\
A^{(3)} &=&3\mu
+2c^{2}-\frac{2}{|c|}p^{(1)}-\frac{8}{3|c|}p^{(2)}-\frac{1}{|c|}
p^{(3)}. \label{BBB}
\end{eqnarray}
We emphasize that the pressure given by (\ref{ppp}) provides the
exact equation of state through iteration with (\ref{BBB}).
The thermodynamics and critical behaviour can thus be worked out in
a straightforward manner in terms of a special polylogarithm
function.

\subsection{Phase diagram in the $\mu-H$ plane}

We first consider quantum phases in the $\mu-H$ plane at low
temperatures. Although there is no quantum phase transition in 1D
many-body systems at finite temperatures, the TLL leads to a
crossover from relativistic dispersion to nonrelativistic dispersion
between different regimes, which may persist at some non-zero
temperatures \cite{Maeda,Erhai}.
The zero temperature phase diagrams for fixed total number of
particles have been explored earlier \cite{Guan2008prl,Angela}.
The phase diagrams in the $\mu-H$ plane from which quantum
criticality and the finite temperature phase diagrams can be mapped
out are investigated here.
At zero temperature, the $\mu-H$ phase diagrams can be worked out
either from the dressed energy equations obtained from the TBA
equations (\ref{TBA}) in the limit $T\to 0$, or by converting the
critical fields in the $H-n$ plane, which were found in
\cite{Guan2008prl}, into the $\mu-H$ plane or directly using the
equation of state (\ref{ppp}) with $T\to 0$.

We first work out the phase diagram for equally-spaced splitting
($H_1=H_2$) at $T=0$ through analyzing the band filling in the
dressed energy equations \cite{Guan2008prl}.
Here we find that the critical field for the phase transition from
the vacuum into the fully trionic phase is
$\mu_c\ge-\frac{2}{3}c^2$.  The critical field for the phase
transition from the fully trionic phase into the mixture of trions
and unpaired fermions is determined by the set of equations
\begin{widetext}
\begin{eqnarray}
\mu_c&\ge&-H_1-\frac{1}{2\pi}\int_{Q_3}^{Q_3}\frac{2|c|}{c^2+\lambda^2}
  \varepsilon_3(\lambda)d\lambda \nonumber\\
  \varepsilon_3(\lambda)&=&3\lambda^2-2c^2-3\mu-\frac{1}{2\pi}\int_{-Q_3}^{Q_3}
  \left[\frac{2|c|}{c^2+(\lambda-\lambda')^2}
+\frac{4|c|}{4c^2+(\lambda-\lambda')^2}\right]\varepsilon_3(\lambda')d\lambda' , \nonumber\\
  {Q_3}^2&=&\frac{2}{3}c^2 +\mu +\frac{1}{6\pi}\int_{-Q_3}^{Q_3} \left[\frac{2|c|}{c^2+\lambda^2}
 +\frac{4|c|}{4c^2+\lambda^2}\right]\varepsilon_3(\lambda)d\lambda.
\label{phase-II}
\end{eqnarray}
\end{widetext}

It seems to be very difficult to get a general expression for
$\mu_c$ from the condition (\ref{phase-II}), except for in the
strong and weak coupling regimes.  Nevertheless, we can extract the
phase boundary by numerical calculation for arbitrary strong
interaction. The critical field for the phase transition from the
vacuum into the fully polarized  phase is given by $\mu_c\ge-H_1$.
The critical field for the  phase transition from the
fully-polarized  phase into the mixed  phase of trions and unpaired
fermions is
\begin{equation}
\mu_c\ge-\frac{2}{3}c^2+\frac{2|c|}{3\pi}
\left[\frac{Q_1^2+c^2}{|c|} \arctan\frac{{Q_1}}{|c|}-{Q_1} \right],
\end{equation}
with $Q_1=\sqrt{\mu+H_1}$. This phase diagram is shown in
Fig.~\ref{fig:mu-h}(a).

The phase boundaries for nonlinear Zeeman splitting are obtained in
a similar fashion. Indeed we find that all zero temperature phase
diagrams are consistent with the $\mu $-$H$ phase diagrams which are
directly plotted from the equation of state (\ref{ppp}) with the
temperature $T=0.001\varepsilon _{b}$, see Fig.~\ref{fig:mu-h}.  For
simplicity, we used $A$, $B$ and $C$ to respectively denote the
phases of unpaired fermions, pairs and trions. The phases $A$+$B$,
$B$+$C$, $A$+$C$ and $A$+$B$+$C$ stand for a mixture of
corresponding phases.

The quantum phase segments in an harmonic trapping potential can
clearly be discerned from the phase diagrams in Fig.~\ref{fig:mu-h}.
The phase diagram in Fig.~\ref{fig:mu-h}(a) is for pure Zeeman
splitting ($H_1=H_2$). The multi-critical point in the phase diagram
in Fig.~\ref{fig:mu-h}(a) is located at $\left(\frac{4\varepsilon
_{b}}{3},-\frac{4\varepsilon _{b}}{3} \right)$ at $T=0$.
It may persist for some non-zero temperatures due to the existence
of TLL phases.
In an harmonic trapping potential, the mixture of trions and
unpaired atoms is  at the centre of the trap, whereas the unpaired
fermions are at the outer wings when the external field
$H>{4\varepsilon_{b}}/{3}$.
However, for $H<{4\varepsilon_{b}}/{3}$ almost the whole cloud is
the trion phase due to a large binding energy of trions.
The mixture of trions and unpaired fermions might lie in a very
narrow strip in the trapping centre.

Quantum phase diagrams for unequally-spaced splittings are very
intriguing. In the phase diagram Fig.~\ref{fig:mu-h}(d) the Zeeman
splitting parameters are $H_2=2H_1$. In this case, the pair phase is
energetically favoured. From the dressed energy equations we can
find that the phase boundaries intersect at
$\left(\frac{10\varepsilon_{b}}{12},-\frac{4\varepsilon _{b}}{3}
\right)$ at $T=0$.
In an harmonic trapping potential, when the external field
$H>\frac{10\varepsilon_{b}}{12}$, the centre of trap is a mixture of
trions and pairs whereas the outer wings are occupied by pairs.
However, for $H<\frac{10\varepsilon_{b}}{12}$, the mixture of trions
and paired fermions lie in a narrow strip in the trapping centre.
The trions occupy the outer wings.

More subtle quantum phases can be tuned through nonlinear Zeeman
splitting, see the phase diagrams in Fig.~\ref{fig:mu-h}(b) and
Fig.~\ref{fig:mu-h}(c), where the phase diagrams for the chemical
potential are shown for the illustrative field values $H_2=1.24H_1$
and $H_2=1.3H_1$.
The mixture of trions, pairs and unpaired fermions can occur in a
certain setting of Zeeman splitting among the three lowest energy
levels.  The intersection points in the phase diagrams can be easily
determined through the equation of state (\ref{ppp}) with such
settings, but it seems to be more difficult to analytically
determine the phase boundaries.  These subtle quantum phases can be
mapped out through the new scheme proposed in \cite{Ho-Zhou} from
experimental data in trapped 1D Fermi gases. In order to understand
the nature of such quantum phases, we turn to the examination of the
specific heat in the $T-H$ plane.

\subsection{Specific heat and entropy}

The thermodynamics of the system (\ref{Ham}) can be analytically
calculated through the equation of state (\ref{ppp}). All
thermodynamic properties then follow analytically through the
general thermodynamic relations. According to the formula for the
specific heat $c_{v}=\left( \partial ^{2}p/\partial
T^{2}\right)_{v}$, the phase diagrams as revealed by $c_{v}$ in the
$T-H$ plane can be easily explored for fixed total density. Here the
specific heat $c_{v}$ is a  function of $T$, $\mu$, $H_{1}$ and
$H_{2}$. Thus the full $c_v$ phase diagram would be four
dimensional. In order to observe the signatures of the TLL, we take
two-dimensional contour plots for the $c_v$ phase diagrams for some
illustrative values of Zeeman splitting associated with
Fig.~\ref{fig:mu-h}.

For pure Zeeman splitting, the gapless phase is described by a
two-component TLL phase under a crossover temperature (lines of
squares in Fig.~\ref{fig:Cv1}) which indicates a deviation from the
linear temperature-dependent specific heat
\begin{equation}
c_v \approx \frac{\pi
T}{3\hbar}\left(\frac{1}{v_{1}}+\frac{1}{v_{3}}\right).
\end{equation}
The trions and unpaired fermions can form an asymmetric
two-component TLL of composite fermions and single atoms for
temperatures below the lines of squares. However, the trion phase
$C$ and unpaired fermions phase $A$ form two different
single-component TLLs which lie below the left and right lines of
triangles, respectively.
In the single-component TLL phase the other states are exponentially
small and thus the system is strongly correlated.

 For unequally spaced Zeeman splitting ($H_{2}=2H_{1}$) the zero
 temperature phase diagram in Fig.~\ref{fig:mu-h}(d) may persist for finite $T$ as
 long as  the excitations are close to the Fermi points of each Fermi sea.
 From the low temperature phase diagram
 Fig.~\ref{fig:Cv2} we see clearly that a two-component TLL of trions and
 pairs remains in the regime $B+C$. The gapless phase is described by a
 two-component TLL phase under a crossover temperature delineated
 by a deviation from the linear temperature-dependent specific
 heat
\begin{equation}
c_v \approx \frac{\pi
T}{3\hbar}\left(\frac{1}{v_{2}}+\frac{1}{v_{3}}\right).
\end{equation}
In this case a TLL of hard-core bosons of composite fermions lies
below the right line of triangles.

For unequally spaced Zeeman splitting ($H_{2}=1.2H_{1}$) the
 three-component TLL ($A+B+C$) and two-component TLL ($A+B$) phases
 may persist within certain regimes in the $T-H$ plane, see the lines of squares in Fig.~\ref{fig:Cv3}.
 Beyond the universal crossover temperatures one of the excitations among the
 states of trions, pairs and unpaired fermions exhibits
 nonrelativistic dispersion.
 In the three-component TLL phase, i.e. where trions, pairs and unpaired fermions coexist, the
 specific heat is given by the linear relation
\begin{equation}
c_v \approx \frac{\pi
T}{3\hbar}\left(\frac{1}{v_{1}}+\frac{1}{v_{2}}+\frac{1}{v_{3}}\right).
\end{equation}
We see clearly that the equation of state (\ref{ppp}) provides a
 precise description of the thermodynamics and critical behaviour of
 composite fermions.

In Fig.~\ref{fig:S}, we demonstrate that the entropy exhibits a peak
as the driving parameter chemical potential varies across a phase
boundary in the $\mu-H$ plane, see Fig.~\ref{fig:mu-h}(c).
The entropy curves are shown in Fig.~\ref{fig:S} for the indicative
values $H_1=1.2 \varepsilon_{b}, \, 1.32 \varepsilon_{b}$ and
$H_1=1.38 \varepsilon_{b}$. In this example, the chemical potential thus varies across the
different phase boundaries in Fig.~\ref{fig:mu-h}(c) at which the quantum phase transitions
occur. The entropy peaks in Fig.~\ref{fig:S} are located in the phases with higher density
of states.

\section{Conclusion}
\label{sec:conclusion}

In conclusion, we have studied the thermodynamics of 1D strongly
attractive three-component fermions in the presence of nonlinear
Zeeman fields via the thermodynamic Bethe ansatz solution.
The pressure and free energy have been analytically
calculated in terms of the chemical potential $\mu$, temperature $T$
and Zeeman fields $H_1$ and $H_2$ for a parameter regime $T\ll
\varepsilon_b, \, \varepsilon_t, \,  H_1, \, H_2$ and $\gamma \gg1$.
Here $\varepsilon_b$ and $ \varepsilon_t$ are the binding energies
for a bound pair and a trion, respectively. This physical regime
covers the presently accessible experimental parameter regime \cite{Hulet}.
The universal thermodynamics of the asymmetric two-component and
three-component TLLs has been identified at low temperatures.
Beyond a certain crossover temperature, at least one of the underlying dispersion
relations for the composite particles is no longer linear and exhibits rich thermal
excitations.

We have derived the equation of state (\ref{ppp}) from
which quantum criticality and quantum phase transitions can be
mapped out.
The equation of state provides the necessary information to describe the quantum regime near quantum critical points.
The scaling functions and critical exponents can be obtained from the equation of state following the approach
for the two-component model \cite{Guan-Ho}.
With regard to the harmonic trapping of three-component fermions, quantum criticality can be mapped out through the
specific heat phase diagram in the $T-\mu$ plane. For example, for
equally-spaced Zeeman splitting with $H_1=1.36 \varepsilon_{b}$,
the critical behaviour of the system can be conceived from the specific
heat phase diagram in the $T-\mu$ plane, see Fig.~\ref{fig:c_v-mu}.
Our results thus open the way for further study of quantum criticality in 1D many-body systems via their exact Bethe ansatz solution.
In this case for systems of
three-component ultracold fermionic atoms.

\acknowledgments

This work is in part supported by NSFC, the Knowledge Innovation
Project of Chinese Academy of Sciences, the National Program for
Basic Research of MOST (China) and the Australian Research Council.
MTB and XWG thank the Institute of Physics, Chinese Academy of
Sciences for kind hospitality during various stages of this work.

\begin{widetext}

\appendix

\section{Derivation of the TBA equations}

For the 1D three-component fermion system we consider, there are
three kinds of states in the system, i.e., unpaired fermions, pairs
and trions. In the thermodynamic limit and at zero temperature,
there are three kinds of quasimomenta solutions to the BA equations
(\ref{BAE}).
These are real $\left\{k_i\right\}$, with $i=1,\ldots, N_1$ for the
unpaired fermions, complex roots $\left\{k_l =\Lambda_{\ell} \pm
\frac{1}{2}\mathrm{i}|c|\right\}$ with $\ell=1,\ldots, N_2$ for
bound pairs and three-body bound states $\left\{k_m =\lambda_m \pm
\mathrm{i}|c|,\, \lambda_m\right\}$ with $m=1,\ldots, N_3$ for
trions.

For finite temperatures, there are also spin strings for spin
rapidities $\Lambda$ and $\lambda$, which are characterized by the
string-hypothesis
\begin{eqnarray}
\Lambda _{j}^{n,\beta} &=& \Lambda
_{j}^{n}-\frac1 2(n+1-2\beta)\mathrm{i}, \\
\lambda _{j}^{n,\beta} &=& \lambda _{j}^{n}-\frac1
2(n+1-2\beta)\mathrm{i}.
\end{eqnarray}
where $n$ is the length of the string, $j$ labels the number of
strings of length $n$, and $\Lambda _{j}^{n}$ and $\lambda _{j}^{n}$
are the real parts of each $\Lambda$ and $\lambda$ string.
At finite temperatures, there are $N_{1}'$ real quasimomenta
$k_{j}$, $N_{2}'$ real  $\Lambda _{j}$ and $ N_{3}'$ real $\lambda
_{j}$. The number of $\Lambda ^{(n)}$-strings is $M_{1n}$ and the
number of the $\lambda ^{(n)}$-strings  is $M_{2n}$. These quantum
numbers satisfy the conditions
\begin{eqnarray}
M_1&=&N_2'+ 2 N_3'+\sum_{n=1}^{\infty}nM_{1n},\\
M_2&=&N_3'+\sum_{n=1}^{\infty}nM_{2n}.
\end{eqnarray}

Substituting these three sets of solutions into the BA equations
(\ref{BAE}) gives
\begin{eqnarray}
\mathrm{e}^{\mathrm{i} k_{j}L}
&=&\prod_{l=1}^{N_{2}^{\prime}}\frac{k_{j}-\Lambda
_{l}+\mathrm{i}|c|/2}{k_{j}-\Lambda _{l}-\mathrm{i}|c|/2}
\prod_{l=1}^{N_{3}^{\prime }}\frac{k_{j}-\lambda _{l}+{ \mathrm{i}
|c|}}{k_{j}-\lambda _{l}-{\mathrm{i} |c|}} \prod_{n=1}^{\infty
}\prod_{l=1}^{M_{1n}}\frac{k_{j}-\Lambda_{l}^{n}+\mathrm{i}n|c|/2}{k_{j}-\Lambda
_{l}^{n}-\mathrm{i} n|c|/2} \label{unpairfinal}
\end{eqnarray}
for unpaired fermions and
\begin{eqnarray}
\mathrm{e}^{2 \mathrm{i}  \Lambda _{j}L} &=&\prod_{l=1,l\neq
j}^{N_{2}^{\prime }}\frac{ \Lambda _{j}-\Lambda _{l}+ \mathrm{i}
|c|} {\Lambda_{j}-\Lambda_{l}- \mathrm{i} |c|}
\left( \frac{k_{j}^{(1)}-\Lambda _{j}+\mathrm{i}|c|/2}{%
k_{j}^{(1)}-\Lambda _{j}
-\mathrm{i}|c|/2}\frac{k_{j}^{(2)}-\Lambda_{j}+\mathrm{i}|c|/2}{k_{j}^{(2)}
-\Lambda_{j}-\mathrm{i}|c|/2}\right)
 \prod_{l=1}^{N_{3}^{\prime }}\frac{\Lambda _{j}-\lambda_{l}+3\mathrm{i}|c|/2}{\Lambda _{j}-\lambda _{l}-3\mathrm{i}|c|/2%
}\frac{\Lambda _{j}-\lambda _{l}+\mathrm{i}|c|/2}{\Lambda_{j}-\lambda _{l}-\mathrm{i}|c|/2}  \notag \\
& & \times \prod_{n=1}^{\infty }\prod_{l=1}^{M_{1n}}\prod_{\beta =1}^{n}\frac{%
\Lambda _{j}-\Lambda _{l}^{n,\beta
}+\mathrm{i}|c|}{\Lambda_{j}-\Lambda _{l}^{n,\beta }-\mathrm{i}|c|}
\label{pair1}
\end{eqnarray}
for paired fermions. Notice that Eq. (\ref{pair1}) has explicit
singularities from the terms in the $()$ bracket.
To overcome this, we write the second term of equation of
(\ref{BAE}) as
\begin{eqnarray}
&&
\prod_{l=1,l\neq j}^{N_{1}^{\prime }}\frac{\Lambda _{j}-k_{l}+%
\mathrm{i}|c|/2}{\Lambda _{j}-k_{l}-\mathrm{i}|c|/2} \left(\frac{
 k_{j}^{(1)}-\Lambda _{j}-\mathrm{i}|c|/2}{k_{j}^{(1)}-\Lambda _{j}+%
\mathrm{i}|c|/2} \frac{k_{j}^{(2)}-\Lambda _{j}-\mathrm{i}|c|/2}{%
k_{j}^{(2)}-\Lambda _{j}+\mathrm{i}|c|/2}\right)  \notag\\
&& \quad = \prod_{n=1}^{\infty
}\prod_{m=1}^{M_{1n}}\prod_{\beta=1}^{n} \frac{\Lambda _{j}-\Lambda
_{m}^{n,\beta }+\mathrm{i}|c|} {\Lambda_{j}-\Lambda_{m}^{n,\beta
}-\mathrm{i}|c|}
 \prod_{n=1}^{\infty }\prod_{m'=1}^{M_{2n}}
\prod_{\alpha=1}^{n}
\frac{\Lambda _{j}-\lambda_{m'}^{n,\alpha }-\mathrm{i}|c|/2}{\Lambda _{j}-\lambda _{m'}^{n,\alpha }+%
\mathrm{i}|c|/2}  \label{pair2}
\end{eqnarray}
which shows the spin flipping of $\Lambda^n$-strings.
Substituting (\ref{pair2}) back to (\ref{pair1}) gives the revised
form
\begin{eqnarray}
\mathrm{e}^{2 \mathrm{i} \Lambda _{j} L}
&=&\prod_{l=1}^{N_{1}^{\prime }}
\frac{\Lambda_{j}-k_{l}+\mathrm{i}|c|/2}{\Lambda_{j}-k_{l}-\mathrm{i}|c|/2}
\prod_{l=1}^{N_{2}^{\prime}} \frac{\Lambda
_{j}-\Lambda_{l}+\mathrm{i}|c|}{\Lambda _{j}-\Lambda
_{l}-\mathrm{i}|c|}
 \prod_{l=1}^{N_{3}^{\prime }} \frac{\Lambda_{j}
-\lambda_{l} + 3\mathrm{i}|c|/2}{\Lambda
_{j}-\lambda_{l}-3\mathrm{i}|c|/2} \frac{\Lambda _{j}
-\lambda_{l}+\mathrm{i}|c|/2}{\Lambda_{j}-\lambda _{l}-\mathrm{i}|c|/2}  \notag \\
&&\times \prod_{n=1}^{\infty }\prod_{l=1}^{M_{2n}}
\frac{\Lambda_{j}-\lambda _{l}^{n}+\mathrm{i}n|c|/2}{\Lambda
_{j}-\lambda_{l}^{n}-\mathrm{i}n|c|/2} \label{pairfinal}
\end{eqnarray}
for pairs without singularities. Similarly, the equation for trions
is
\begin{eqnarray}
\mathrm{e}^{3 \mathrm{i} \lambda _{j} L} &=&
\prod_{l=1}^{N_{1}^{\prime
}}\frac{\lambda_{j}-k_{l}+\mathrm{i}|c|}{\lambda
_{j}-k_{l}-\mathrm{i}|c|} \prod_{l=1}^{N_{2}^{\prime }}
\frac{\lambda _{j}-\Lambda_{l} +\mathrm{i}|c|/2}{\lambda _{j} -\Lambda _{l} -\mathrm{i}|c|/2}%
\frac{\lambda _{j} -\Lambda _{l} +3\mathrm{i}|c|/2}{\lambda_{j} -\Lambda _{l} -3\mathrm{i}|c|/2}  \notag \\
&& \times \prod_{l=1}^{N_{3}^{\prime }}\frac{\lambda _{j}
-\lambda_{l} +\mathrm{i}|c|}{\lambda _{j} -\lambda _{l}
-\mathrm{i}|c|}\frac{\lambda _{j} -\lambda _{l} +2\mathrm{i}|c|}
{\lambda _{j} -\lambda _{l} -2\mathrm{i}|c|}. \label{trionfinal}
\end{eqnarray}

The BA equations for the spin parts are
\begin{eqnarray}
\prod_{l=1}^{N_{1}^{\prime }}\frac{\Lambda _{j}^{m}-k_{l}
-\mathrm{i}m|c|/2}{\Lambda _{j}^{m}-k_{l} +\mathrm{i}m|c|/2}
&=&-\prod_{n=1}^{\infty }\prod_{l=1}^{M_{1n}}\prod_{\beta =1}^{n}\frac{%
\Lambda _{j}^{m}-\Lambda _{l}^{n}+(m+n+2-2\beta
)\mathrm{i}|c|/2}{\Lambda_{j}^{m} -\Lambda
_{l}^{n}-(m+n+2-2\beta)\mathrm{i}|c|/2}
\,\, \frac{\Lambda _{j}^{m}-\Lambda _{l}^{n}+(m+n-2\beta )\mathrm{i}|c|/2}{%
\Lambda _{j}^{m}-\Lambda _{l}^{n}-(m+n-2\beta )\mathrm{i}|c|/2}  \notag \\
&& \times \prod_{n=1}^{\infty }\prod_{l=1}^{M_{2n}}\prod_{\beta =1}^{n}\frac{%
\Lambda _{j}^{m}-\lambda _{l}^{n}+(m+n+1-2\beta
)\mathrm{i}|c|/2}{\Lambda _{j}^{m}-\lambda _{l}^{n}-(m+n+1-2\beta
)\mathrm{i}|c|/2},
\end{eqnarray}
\begin{eqnarray}
\prod_{l=1}^{N_{2}^{\prime }}\frac{\lambda _{j}^{m}-\Lambda _{l} +%
\mathrm{i}m|c|/2}{\lambda _{j}^{m}-\Lambda _{l} -\mathrm{i}m|c|/2}
&=&-\prod_{n=1}^{\infty }\prod_{l=1}^{M_{2n}}\prod_{\beta =1}^{n}\frac{%
\lambda _{j}^{m}-\lambda _{l}^{n}+(m+n+2-2\beta
)\mathrm{i}|c|/2}{\lambda _{j}^{m}-\lambda _{l}^{n}-(m+n+2-2\beta
)\mathrm{i}|c|/2}
\,\, \frac{\lambda _{j}^{m}-\lambda _{l}^{n}+(m+n-2\beta )\mathrm{i}|c|/2}{%
\lambda _{j}^{m}-\lambda _{l}^{n}-(m+n-2\beta )\mathrm{i}|c|/2}  \notag \\
&& \times \prod_{n=1}^{\infty }\prod_{l=1}^{M_{1n}}\prod_{\beta =1}^{n}\frac{%
\lambda _{j}^{m}-\Lambda _{l}^{n}+(m+n+1-2\beta
)\mathrm{i}|c|/2}{\lambda _{j}^{m}-\Lambda _{l}^{n}-(m+n+1-2\beta
)\mathrm{i}|c|/2}.
\end{eqnarray}

Defining the function $\theta (x)=2\arctan x $ and taking the
logarithm on both sides of the above equations  gives
\begin{eqnarray}
k_{j}L&=&2\pi I_{j}+\sum_{l=1}^{N_{2}^{\prime }}\theta
(\frac{k_{j}-\Lambda_{l}}{|c^{\prime}|})+\sum_{l=1}^{N_{3}^{\prime}}\theta
(\frac{k_{j}-\lambda_{l}}{2|c^{\prime}|}) +
\sum_{n=1}^{\infty}\sum_{l=1}^{M_{1n}}\theta
(\frac{k_{j}-\Lambda_{l}^{n}}{n|c^{\prime}|}), \\
2\Lambda _{j}L&=&2\pi J_{j}+\sum_{l=1}^{N_{1}^{\prime }}\theta
(\frac{ \Lambda
_{j}-k_{l}}{|c^{\prime}|})+\sum_{l=1}^{N_{2}^{\prime}}\theta
(\frac{\Lambda _{j}-\Lambda _{l}}{2|c^{\prime}|})
 + \sum_{l=1}^{N_{3}^{\prime }}\left[\theta (\frac{\Lambda_{j}-\lambda _{l}}{ 3|c^{\prime}|})
 +\theta (\frac{\Lambda_{j}-\lambda _{l}}{|c^{\prime}|})\right]  \notag\\
 && + \sum_{n=1}^{\infty }\sum_{l=1}^{M_{2n}}
 \theta (\frac{\Lambda_{j}-\lambda_{l}^{n}}{n|c^{\prime}|}), \\
3\lambda _{j}L&=&2\pi K_{j}+\sum_{l=1}^{N_{1}^{\prime }}\theta
(\frac{\lambda _{j}-k_{l}}{2|c^{\prime}|})
 + \sum_{l=1}^{N_{2}^{\prime }}\left[\theta (\frac{\lambda_{j}-\Lambda _{l}}{ |c^{\prime}|})
 +\theta (\frac{\lambda_{j}-\Lambda_{l}}{3|c^{\prime}|})\right]  \notag \\
&& + \sum_{l=1}^{N_{3}^{\prime }}\left[\theta
(\frac{\lambda_{j}-\lambda _{l}}{ 2|c^{\prime}|}) +\theta
(\frac{\lambda_{j}-\lambda _{l}}{4|c^{\prime}|})\right],
\end{eqnarray}
\begin{eqnarray}
\sum_{l=1}^{N_{1}^{\prime }}\theta
(\frac{\Lambda_{j}^{m}-k_{l}^{n}}{ m|c^{\prime}|})&=&2\pi
I_{j}^{(n)}+\sum_{n=1}^{\infty
}\sum_{l=1}^{M_{1n}}\Theta_{mn}(\frac{\Lambda _{j}^{m}-\Lambda
_{l}^{n}}{ |c^{\prime}|}) - \sum_{n=1}^{\infty
}\sum_{l=1}^{M_{2n}}\Xi _{mn}(\frac{\Lambda_{j}^{m}-\lambda
_{l}^{n}}{|c^{\prime}|}),
\end{eqnarray}
\begin{eqnarray}
\sum_{l=1}^{N_{2}^{\prime }}\theta (\frac{\lambda
_{j}^{m}-\Lambda_{l}^{n} }{m|c^{\prime}|})&=&2\pi J_{j}^{(n)}
+\sum_{n=1}^{\infty}\sum_{l=1}^{M_{2n}}\Theta _{mn}(\frac{\lambda
_{j}^{m}-\lambda_{l}^{n}}{|c^{\prime}|})
 - \sum_{n=1}^{\infty }\sum_{l=1}^{M_{1n}}\Xi _{mn}(\frac{\lambda_{j}^{m}-\Lambda _{l}^{n}}{|c^{\prime}|}).
\end{eqnarray}
Here $c' = c/2$ with $I_j$, $J_j$, $K_j$, $I_j^{(n)}$, $J_j^{(n)}$
integers or half-odd-integers depending on the quantum numbers.

The functions $\Theta_{mn}$ and $\Xi_{mn}$ are defined by
\begin{eqnarray}
\Theta_{mn} &=&\left\{
\begin{array}{l}
\theta_{m+n}+2\theta_{m+n-2}+\cdots 2\theta_{|m-n|+2}+\theta_{|m-n|}, \quad \mathrm{for} \,\, m\neq n,\nonumber \\
2\theta_{2}+2\theta_{4}+\cdots +2\theta_{2n-2}+\theta_{2n}, \quad
\mathrm{for}  \,\, m=n,
\end{array}
\right.  \\
\Xi_{mn} &=&\left\{
\begin{array}{l}
\theta_{m+n-1}+\theta_{m+n-3}+\cdots
\theta_{|m-n|+3}+\theta_{|m-n|+1}, \quad
\mathrm{for} \,\, m\neq n,\nonumber \\
\theta_{1}+\theta_{3}+\cdots +\theta_{2n-3} +\theta_{2n-1}, \quad
\mathrm{for} \,\, m=n.
\end{array}
\right.
\end{eqnarray}

Finally we define the functions
\begin{eqnarray}
h^{\prime }(k) &=&k L-\sum_{l=1}^{N_{2}^{\prime }}\theta
(\frac{k-\Lambda _{l}
}{|c^{\prime}|})-\sum_{l=1}^{N_{3}^{\prime}}\theta (\frac{k-\lambda
_{l}}{2|c^{\prime}|}) - \sum_{n=1}^{\infty
}\sum_{l=1}^{M_{1n}}\theta (\frac{k-\Lambda_{l}^{n}}{
n|c^{\prime}|}),
\end{eqnarray}
\begin{eqnarray}
j^{\prime }(\Lambda ) &=&2\Lambda
L-\sum_{l=1}^{N_{1}^{\prime}}\theta ( \frac{\Lambda
-k_{l}}{|c^{\prime}|})-\sum_{l=1}^{N_{2}^{\prime }}\theta (
\frac{\Lambda -\Lambda _{l}}{2|c^{\prime}|})
 - \sum_{l=1}^{N_{3}^{\prime }}\left[\theta (\frac{\Lambda -\lambda_{l}}{ 3|c^{\prime}|})+\theta (\frac{\Lambda
-\lambda_{l}}{|c^{\prime}|}) \right]  - \sum_{n=1}^{\infty
}\sum_{l=1}^{M_{2n}}\theta (\frac{\Lambda
-\lambda_{l}^{n}}{n|c^{\prime}|}),
\end{eqnarray}
\begin{eqnarray}
k^{\prime }\left( \lambda \right)&=&3\lambda
L-\sum_{l=1}^{N_{1}^{\prime}}\theta (\frac{\lambda
-k_{l}}{2|c^{\prime}|})
 - \sum_{l=1}^{N_{2}^{\prime }}
 \left[\theta (\frac{\lambda -\Lambda_{l}}{ |c^{\prime}|})+\theta (\frac{\lambda -\Lambda _{l}}{3|c^{\prime}|})\right]
 - \sum_{l=1}^{N_{3}^{\prime }}\left[\theta (\frac{\lambda -\lambda_{l}}{ 2|c^{\prime}|})+\theta (\frac{\lambda
-\lambda_{l}}{4|c^{\prime}|})\right],
\end{eqnarray}
and
\begin{eqnarray}
j_{m}(\Lambda ^{m}) &=&\sum_{l=1}^{N_{1}^{\prime }}\theta
(\frac{\Lambda ^{m}-k_{l}^{n}}{m|c^{\prime}|})-\sum_{n=1}^{\infty
}\sum_{l=1}^{M_{1n}}\Theta_{mn}(\frac{\Lambda ^{m}-\Lambda
_{l}^{n}}{|c^{\prime}|}) + \sum_{n=1}^{\infty
}\sum_{l=1}^{M_{2n}}\Xi _{mn}(\frac{\Lambda
^{m}-\lambda _{l}^{n}}{|c^{\prime}|}), \\
k_{m}(\lambda ^{m}) &=&\sum_{l=1}^{N_{1}^{\prime }}\theta
(\frac{\lambda ^{m}-\Lambda _{l}^{n}}{m|c^{\prime}|})
-\sum_{n=1}^{\infty }\sum_{l=1}^{M_{2n}}\Theta _{mn}(\frac{\lambda
^{m}-\lambda _{l}^{n}}{ |c^{\prime}|})
 + \sum_{n=1}^{\infty }\sum_{l=1}^{M_{1n}}\Xi _{mn}(\frac{\lambda
^{m}-\Lambda _{l}^{n}}{|c^{\prime}|}).
\end{eqnarray}

In the thermodynamic limit, we then define
\begin{eqnarray}
\frac{dh^{\prime }(k)}{dk} &=&2\pi (\rho _{1}(k)+\rho _{1}^{h}(k)), \\
\frac{dj^{\prime }(\Lambda )}{d\Lambda } &=&2\pi (\rho _{2}(\Lambda)+\rho_{2}^{h}(\Lambda )), \\
\frac{dk^{\prime }\left( \lambda \right) }{d k} &=&2\pi
(\rho_{3}\left(
\lambda \right) +\rho _{3}^{h}\left( \lambda \right) ), \\
\frac{dj_{n}(\Lambda ^{n})}{d\Lambda ^{n}} &=&2\pi
(\sigma_{n}(\Lambda
^{n})+\sigma _{n}^{h}(\Lambda ^{n})), \\
\frac{dk_{n}(\lambda ^{n})}{d\lambda ^{n}} &=&2\pi (\tau_{n}(\lambda
^{n})+\tau _{n}^{h}(\lambda ^{n})).
\end{eqnarray}
where $\rho_{i}$  and $\rho_{i}^h$ for $i=1,2,3$ are particle and
hole densities in $k$-space, and $\sigma _{n}$, $\sigma_{n}^{h}$ and
$\tau_{n}$, $\tau_{n}^{h}$ are particle densities and hole densities
for strings with length $n$ in $\Lambda$-space and $\lambda$-space.


Thus we have the integral equations
\begin{eqnarray}
\frac{1}{2\pi} &=& \rho_1 + \rho_1^h +a_1*\rho_2 + a_2*\rho_3  + \sum_n a_n*\sigma_n, \\
\frac{1}{\pi} &=& \rho_2 + \rho_2^h+a_1*\rho_1 + a_2*\rho_2  + \, (a_1+a_3)*\rho_3+ \sum_n a_n*\tau_n, \\
\frac{3}{2\pi} &=& \rho_3 + \rho_3^h+ a_2*\rho_1 + (a_1+a_3)*\rho_2
+ \, (a_2+a_4)*\rho_3,
\end{eqnarray}
for the particle and hole densities and
\begin{eqnarray}
a_n*\rho_1&=&\sigma_n + \sigma_n^h +\sum_m T_{mn}*\sigma_m
  - \, \sum_m S_{mn}*\tau_m, \\
a_n*\rho_2&=&\tau_n + \tau_n^h +\sum_m T_{mn}*\tau_m   - \, \sum_m
S_{mn}*\sigma_m.
\end{eqnarray}
where the functions $T_{mn}$ and $S_{mn}$ are defined as
\begin{eqnarray}
T_{mn} &=&\left\{
\begin{array}{l}
a_{m+n}+2a_{m+n-2}+\cdots 2a_{|m-n|+2}+a_{|m-n|}, \quad \mathrm{for} \,\,  m\neq n,\nonumber \\
2a_{2}+2a_{4}+\cdots +2a_{2n-2}+a_{2n}, \quad \mathrm{for}  \,\,
m=n,
\end{array}
\right.
\end{eqnarray}
and
\begin{eqnarray}
S_{mn} &=&\left\{
\begin{array}{l}
a_{m+n-1}+a_{m+n-3}+\cdots a_{|m-n|+3}+a_{|m-n|+1},\quad
\mathrm{for}  \,\,  m\neq n,\nonumber \\
a_{1}+a_{3}+\cdots +a_{2n-3}+a_{2n-1}, \quad \mathrm{for}  \,\, m=n.
\end{array}
\right. \label{T-S}
\end{eqnarray}

The energy per unit length can now be written as
\begin{eqnarray}
E/L &=&\int k^{2}\rho _{1}(k)dk+\int (2k^{2}-\frac{c^{2}}{2})\rho
_{2}(k)dk  + \int (3k^{2}-2c^{2})\rho _{3}(k)dk.
\end{eqnarray}
The total particle number and magnetic numbers are
\begin{eqnarray}
N/L&=& \int (\rho _{1}(k)+2\rho _{2}(k)+3\rho _{3}(k))dk,\\
M_{1}/L &=&\int \rho _{2}(k)dk+2\int \rho _{3}(k)dk
+ \, \sum_{n}n\int \sigma _{n}(\Lambda ^{n}) d\Lambda ^{n}, \\
M_{2}/L &=&\int \rho _{3}(k)dk+\sum_{n}n\int \tau _{n}(\lambda ^{n})
d\lambda ^{n}.
\end{eqnarray}
The entropy per unit length  is
\begin{eqnarray}
S/L &=&\int ((\rho _{1}+\rho _{1}^{h})\ln (\rho
_{1}+\rho_{1}^{h})-\rho_{1}\ln \rho _{1}-\rho _{1}^{h}\ln \rho
_{1}^{h})dk + \int ((\rho _{2}+\rho _{2}^{h})\ln (\rho _{2}+\rho
_{2}^{h})-\rho_{2}\ln
\rho _{2}-\rho _{2}^{h}\ln \rho _{2}^{h})dk  \notag \\
&&+\int ((\rho _{3}+\rho _{3}^{h})\ln (\rho _{3}+\rho _{3}^{h})-\rho_{3}\ln \rho _{3}-\rho _{3}^{h}\ln \rho _{3}^{h})dk  \notag\\
&&+ \sum_{n}\int ((\sigma _{n}+\sigma _{n}^{h})\ln
(\sigma_{n}+\sigma_{n}^{h})-\sigma _{n}\ln \rho _{1}\sigma _{n}  -\, \sigma _{n}^{h}\ln \sigma _{n}^{h})dk  \notag\\
&&+ \sum_{n}\int ((\sigma _{n}+\sigma _{n}^{h})\ln
(\sigma_{n}+\sigma_{n}^{h})-\sigma _{n}\ln \rho _{1}\sigma _{n}   -
\, \sigma _{n}^{h}\ln \sigma _{n}^{h})dk,
\end{eqnarray}
where $\ln n! \approx n \ln n$ has been used.

The Gibbs energy (\ref{Gibbs}) per unit length is
\begin{eqnarray}
G/L &=&\int k^{2}\rho _{1}dk+\int (2k^{2}-\frac{c^{2}}{2})\rho_{2}dk
 + \int (3k^{2}-2c^{2})\rho _{3}dk
 -\mu \int (\rho _{1}+2\rho_{2}+3\rho_{3})dk  \notag\\
 &&
 - H_{1}\int (\rho _{1}-2\sum_{n}n\sigma _{n}+\sum_{n}n\tau _{n})dk
 - H_{2}\int (\rho _{2}+\sum_{n}n\sigma _{n}-2\sum_{n}n\tau_{n})dk
- TS. \label{freeG}
\end{eqnarray}
Finally, the TBA equations follow by taking the variation of
equation (\ref{freeG}) and setting it equal to zero, i.e. $\delta
G/L =0$.
In this way
\begin{eqnarray}
\ln \eta _{1} &=& (k^{2}-\mu -H_{1})/{T} \, + a_{1}\ast \ln
(1+\eta_{2}^{-1})
 + \, a_{2} \ast \ln (1+\eta _{3}^{-1})
- \, \sum_{n=1}^\infty a_{n}\ast \ln(1+\xi _{n}^{-1}),
\label{TBA01} \\
\ln \eta _{2} &=& ({2\Lambda ^{2} - \frac12{c^{2}} - 2\mu
-H_{2}})/{T} +a_{1}\ast \ln (1+\eta _{1}^{-1})
 + \, a_{2}\ast \ln (1+\eta _{2}^{-1})+(a_{1}+a_{3})\ast \ln (1+\eta_{3}^{-1}) \notag \\
&& - \, \sum_{n=1}^\infty a_{n}\ast \ln (1+\zeta _{n}^{-1}),  \label{TBA02} \\
\ln \eta _{3} &=&\frac{3\lambda ^{2}-2{c^{2}}-3\mu }{T}+a_{2}\ast
\ln(1+\eta _{1}^{-1})
 + \, (a_{1}+a_{3})\ast \ln (1+\eta _{2}^{-1})
 + \, (a_{2}+a_{4})\ast \ln (1+\eta _{3}^{-1})  \label{TBA03}
\end{eqnarray}
and
\begin{eqnarray}
\ln \xi _{n} &=& {n(2H_{1}-H_{2})}/{T} \, +a_{n}\ast \ln
(1+e^{-{\varepsilon _{1}}/{T}}) + \, \sum_{m}T_{mn}\ast \ln (1+\xi
_{m}^{-1})
 - \, \sum_{m}S_{mn}\ast \ln (1+\zeta _{m}^{-1}),  \label{TBA04} \\
\ln \zeta _{n} &=& {n(2H_{2}-H_{1})}/{T} \,  +a_{n}\ast \ln
(1+e^{-{\varepsilon_{2}}/{T}})
 + \, \sum_{m}T_{mn}\ast \ln (1+\zeta _{m}^{-1})
- \, \sum_{m}S_{mn}\ast \ln (1+\xi _{m}^{-1}),  \label{TBA05}
\end{eqnarray}
in which we define $\eta _{i}=\rho_{i}^{h}/\rho_{i}$ ($i=1,2,3$),
$\xi_{n}=\sigma_{n}^{h}/\sigma_{n}$ and
$\zeta_{n}=\tau_{n}^{h}/\tau_{n}$.
The TBA equations (\ref{TBA01})-(\ref{TBA05}) can be written in the
form (\ref{TBA}) with $\varepsilon_{i}=T\ln \eta_{a}$ ($a=1,2,3$).

\section{Details of the Sommerfeld expansion}

In this Appendix, we show that accurate expressions for $\mu $ and
$p$ can be obtained by iteration after Sommerfeld expansion.
We first use equation (\ref{AAA}) to rewrite the pressure
(\ref{p_gamma}) in the form
\begin{eqnarray}
&&p^{(1)} \approx \frac{2}{3\pi }(\mu ^{(1)})^{{3}/{2}} \left(
1+\frac{\pi ^{2}}{8}\left(\frac{T}{A^{(1)}}\right)^{2}\right)
  \left( 1-\frac{2p^{(2)}}{|c|\mu ^{(1)}}-\frac{2p^{(3)}}{3|c|\mu ^{(1)}}\right)^{{3}/{2}},
\label{aa1} \\
&&p^{(2)} \approx \frac{2\sqrt{2}}{3\pi }(2\mu ^{(2)})^{{3}/{2}}
\left( 1+\frac{\pi ^{2}}{8}\left(\frac{T}{A^{(2)}}\right)^{2}\right)
   \left( 1-\frac{2p^{(1)}}{|c|\mu ^{(2)}}-\frac{p^{(2)}}{2|c|\mu ^{(2)}}-
\frac{8p^{(3)}}{9|c|\mu ^{(2)}}\right) ^{{3}/{2}},
\label{aa2} \\
&&p^{(3)} \approx \frac{2\sqrt{3}}{3\pi }(3\mu ^{(3)})^{{3}/{2}}
\left( 1+\frac{\pi ^{2}}{8}
\left(\frac{T}{A^{(3)}}\right)^{2}\right)     \left(
1-\frac{2p^{(1)}}{3|c|\mu ^{(3)}}-\frac{8p^{(2)}}{9|c|\mu ^{(3)}}-
\frac{p^{(3)}}{3|c|\mu ^{(3)}}\right) ^{{3}/{2}}. \label{aa3}
\end{eqnarray}

We can extract explicit analytic results appropriate for the strong
coupling regime $|c|\gg 1$ from these equations by iteration and
neglecting higher order terms.
Making use of equations (\ref{n}) and $n=n_{1}+2 n_{2}+3 n_{3}$, we
obtain expressions for $n$, $n_{1}$, $n_{2}$ and $n_{3}$. After a
lengthy calculation, we then obtain $\mu ^{(1)}$, $\mu ^{(2)}$ and
$\mu ^{(3)}$ in terms of $n_{1}$, $ n_{2}$ and $n_{3}$, namely
\begin{eqnarray}
\mu ^{(1)}&\approx& n_{1}^{2}{\pi ^{2}}\left[ 1+\frac{16}{3|c|\pi
}\frac{(\mu^{(2)})^{{3}/{2}}}{\mu ^{(1)}}\left( 1+\frac{\pi^{2}}{8}\left(\frac{T}{A^{(2)}}\right)^{2}\right)
+\frac{4}{|c|\pi }\frac{(\mu ^{(3)})^{{3}/{2}}}{\mu^{(1)}}\left(1+\frac{\pi ^{2}}{8}\left(\frac{T}{A^{(3)}}\right)^{2}\right) \right. \notag\\
&&\left. +\frac{16}{|c|}\frac{\sqrt{\mu ^{(2)}}}{\pi }\left(
1-\frac{\pi ^{2}}{24}\left(\frac{T}{A^{(2)}}\right)^{2}\right)
+\frac{12}{|c|}\frac{\sqrt{\mu ^{(3)}}}{\pi }\left(1-\frac{\pi
^{2}}{24}\left(\frac{T}{A^{(3)}}\right)^{2}\right) \right] \left(
1+\frac{\pi ^{2}}{12}\left(\frac{T}{A^{(1)}}\right)^{2}\right),
\label{mumu1}\\
 \mu ^{(2)}& \approx& \frac{n_{2}^{2}}{4}{\pi
^{2}}\left[ 1+\frac{4}{3|c|\pi}\frac{ (\mu ^{(1)})^{{3}/{2}}}{\mu
^{(2)}} \left(
1+\frac{\pi^{2}}{8}\left(\frac{T}{A^{(1)}}\right)^{2}\right)+\frac{16}{3|c|\pi
}\sqrt{\mu ^{(2)}} \right.  \notag \\
&&\left.  +\frac{16}{3|c|\pi }\frac{(\mu ^{(3)})^{{3}/{2}}}{\mu
^{(2)}} \left( 1+\frac{\pi
^{2}}{8}\left(\frac{T}{A^{(3)}}\right)^{2}\right)
+\frac{4}{|c|}\frac{\sqrt{\mu ^{(1)}}}{\pi }\left(
1-\frac{\pi ^{2}}{24}\left(\frac{T}{A^{(1)}}\right)^{2}\right) \right.  \notag \\
&&\left. +\frac{16}{|c|}\frac{\sqrt{\mu ^{(3)}}}{\pi }\left(
1-\frac{\pi ^{2}}{24}\left(\frac{T}{A^{(3)}}\right)^{2}\right)
\right] \left( 1+\frac{\pi
^{2}}{12}\left(\frac{T}{A^{(2)}}\right)^{2}\right),
\label{mumu2}\\
\mu ^{(3)}&\approx& \frac{n_{3}^{2}}{9}{\pi ^{2}}\left[
1+\frac{4}{9|c|\pi }\frac{ (\mu ^{(1)})^{{3}/{2}}}{\mu ^{(3)}}\left(
1+\frac{\pi^{2}}{8}\left(\frac{T}{A^{(1)}}\right)^{2}\right)
+\frac{64}{27|c|\pi }\frac{(\mu ^{(2)})^{{3}/{2}}}{\mu^{(3)}}
\left( 1+\frac{\pi ^{2}}{8}\left(\frac{T}{A^{(2)}}\right)^{2}\right) \right.  \notag \\
&&\left. +\frac{8}{|c|\pi}\sqrt{\mu ^{(3)}}
+\frac{4}{3|c|}\frac{\sqrt{\mu ^{(1)}}}{\pi }\left( 1-\frac{\pi^{2}}{24} \left(\frac{T}{A^{(1)}} \right)^{2}\right)  +\frac{64}{9|c|}\frac{\sqrt{\mu ^{(2)}}}{\pi }\left(1-\frac{\pi ^{2}}{24} \left(\frac{T}{A^{(2)}} \right)^{2}\right) \right]  \notag \\
&&\times \left( 1+\frac{\pi ^{2}}{12}
\left(\frac{T}{A^{(2)}}\right)^{2}\right). \label{mumu3}
\end{eqnarray}

Substituting (\ref{mumu1})-(\ref{mumu3}) and (\ref{p_gamma}) into
(\ref{AAA}) and keeping terms to  order ${1}/{|c|}$, gives the
explicit form for $A^{(a )}$
\begin{eqnarray}
A^{(1)} &\approx &n_{1}^{2}{\pi ^{2}}\left(
1+\frac{8}{|c|}n_{2}+\frac{4}{|c|}
n_{3}\right) , \\
A^{(2)} &\approx &\frac{n_{2}^{2}}{2}{\pi ^{2}}\left(
1+\frac{4}{|c|}n_{1}+
\frac{2}{|c|}n_{2}+\frac{16}{3|c|}n_{3}\right) , \\
A^{(3)} &\approx &\frac{n_{3}^{2}}{3}{\pi ^{2}}\left(
1+\frac{4}{3|c|}n_{1}+
\frac{32}{9|c|}n_{2}+\frac{2}{|c|}n_{3}\right).
\end{eqnarray}
The explicit form for $\mu ^{(a)}$ without $A^{(a)}$ terms can also
be obtained  as
\begin{eqnarray}
\mu^{(1)} &\approx&n_{1}^{2}{\pi ^{2}}\left\{\frac{\pi^{2}}{6}
\left(\frac{T}{n_{1}^{2}{ \pi ^{2}}}\right)^{2} +
\frac{2}{3|c|}\frac{n_{2}^{3}}{n_{1}^{2}}\left[ 1+\frac{\pi ^{2}}{4}
\left (\frac{T}{\frac{n_{2}^{2}}{2}\pi ^{2}} \right)^{2}\right]+
\frac{4}{27|c|}\frac{n_{3}^{3}}{n_{1}^{2}} \left[
1+\frac{\pi^{2}}{4}
\left(\frac{T}{\frac{n_{3}^{2}}{3}\pi ^{2}} \right)^{2}\right] \right.  \notag \\
&&+\left. \left[1 + \frac{8}{|c|}n_{2} + \frac{4}{|c|}n_{3}\right]
\left[ 1-\frac{\pi ^{2}}{12} \left(\frac{T}{n_{1}^{2}{ \pi^{2}}}
\right)^{2}\right] \right\},  \label{mu_1}
\end{eqnarray}
\begin{eqnarray}
\mu^{(2)} & \approx& \frac{n_{2}^{2}}{4}\pi
^{2}\left\{\frac{\pi^{2}}{6} \left(\frac{T}{\frac{ n_{2}^{2}}{2}}
\right)^{2}+ \frac{16}{3|c|}\frac{n_{1}^{3}}{n_{2}^{2}} \left[
1+\frac{\pi^{2}}{4} \left(\frac{T}{n_{1}^{2}{\pi ^{2}}}
\right)^{2}\right] + \frac{64}{81|c|}\frac{n_{3}^{3}}{n_{2}^{2}}
\left[
1+\frac{\pi ^{2}}{4 }\left( \frac{T}{\frac{n_{3}^{2}}{3}}\right) ^{2}\right]+\frac{8}{3|c|}n_2 \right.  \notag \\
&& + \left. \left[1+ \frac{4}{|c|}n_{1} +
\frac{16}{3|c|}n_{3}\right] \left[ 1-\frac{\pi^{2}}{12}
\left(\frac{T}{\frac{ n_{2}^{2}}{2}} \right)^{2}\right] \right\},
\label{mu_2}
\end{eqnarray}
\begin{eqnarray}
\mu^{(3)} &\approx&
\frac{n_{3}^{2}}{9}\pi^{2}\left\{\frac{\pi^{2}}{6}
\left(\frac{T}{\frac{ n_{3}^{2}}{3}{\pi ^{2}}}\right)^{2}+
\frac{4}{|c|}\frac{n_{1}^{3}}{n_{3}^{2}} \left[ 1+\frac{\pi ^{2}}{4}
\left( \frac{T}{n_{1}^{2}{\pi ^{2}}} \right)^{2} \right] +
\frac{8}{3|c|}\frac{n_{2}^{3}}{n_{3}^{2}}
\left[1+\frac{\pi ^{2}}{4}\left(\frac{T}{\frac{n_{2}^{2}}{2}}\right)^{2}\right]+\frac{8}{3|c|}n_3 \right.  \notag \\
&& + \left. \left[1+ \frac{4}{3|c|}n_{1} +
\frac{32}{9|c|}n_{2}\right] \left[ 1+\frac{\pi
^{2}}{12}\left(\frac{T}{\frac{
n_{3}^{2}}{3}{\pi^{2}}}\right)^{2}\right] \right\}. \label{mu_3}
\end{eqnarray}
Substitute all of these equations back into (\ref{aa1})-(\ref{aa3})
and neglecting higher order terms gives the pressure $p$ given in
(\ref{p_se}).

\section{Dressed energy equations at low temperature}

For the strongly attractive regime, the TBA equations (\ref{TBA})
can be expanded as
\begin{eqnarray}
\varepsilon _{1}(k) &\approx&k^{2}-\mu
-H_{1}+\frac{2}{|c|}p^{(2)}+\frac{2}{3|c|} p^{(3)}   - \,
T\sum_{n}\int a_{n}(\frac{2k}{c}-k^{\prime })\ln
(1+\xi^{-1})dk^{\prime},
\label{TBAlc1} \\
\varepsilon _{2}(k) &\approx&2k^{2}-2\mu
-\frac{c^{2}}{2}-H_{2}+\frac{4}{|c|}p^{(1)}+
\frac{1}{|c|}p^{(2)}+\frac{16}{9|c|}p^{(3)}
 - \,T\sum_{n}\int a_{n}(\frac{2k}{c}-k^{\prime })\ln (1+\zeta^{-1})dk^{\prime },  \label{TBAlc2} \\
\varepsilon _{3}(k) &\approx&3k^{2}-3\mu
-2c^{2}+\frac{2}{|c|}p^{(1)}+\frac{8}{3|c|}
p^{(2)}+\frac{1}{|c|}p^{(3)}, \label{TBAlc3}
\end{eqnarray}
with
\begin{eqnarray}
\ln \xi _{n}(\Lambda ) &=&\frac{n(2H_{1}-H_{2})}{T}+\frac{2\pi
J_{1}}{T} a_{n}(\Lambda )
 + \, \sum_{m}T_{mn}\ast \ln (1+\xi _{m}^{-1})
 - \, \sum_{m}S_{mn}\ast \ln (1+\zeta _{m}^{-1}),  \label{TBAlc4} \\
\ln \zeta _{n}(\lambda ) &=&\frac{n(2H_{2}-H_{1})}{T}+\frac{2\pi
J_{2}}{T} a_{n}(\Lambda )
 + \, \sum_{m}T_{mn}\ast \ln (1+\zeta _{m}^{-1})
 - \sum_{m}S_{mn}\ast \ln (1+\xi _{m}^{-1}).  \label{TBAlc5}
\end{eqnarray}
Here $J_{1}=2 p_{1}/|c|$, $J_{2}=p_{2}/|c|$ and the function
$a_{n}(k)$ has the new form
\begin{equation}
a_{n}(k)=\frac{1}{2\pi }\frac{2n}{k^{2}+n^{2}}.
\end{equation}

Compared with equation (\ref{epsilon3}), the last terms of
(\ref{TBAlc1}) and (\ref{TBAlc2}) are string terms for spin waves of
unpaired fermions and pairs, respectively.
From $\ln \xi $ and $\ln \zeta $, we have
\begin{eqnarray}
\xi_{n}(\Lambda ) &\approx& \mathrm{e}^{{n(2H_{1}-H_{2})}/{T}}
\mathrm{e}^{{2\pi J_{1}a_{1}(\Lambda )}/{T}}
  \, \mathrm{e}^{\sum_{m}T_{mn}\ast \xi _{m}^{-1}}  \mathrm{e}^{-\sum_{m}S_{mn}\ast \zeta_{m}^{-1}}, \\
\zeta_{n}(\lambda ) &\approx&  \mathrm{e}^{{n(2H_{2}-H_{1})}/{T}}
\mathrm{e}^{{2\pi J_{2} a_{2}(\lambda )}/{T}}
 \, \mathrm{e}^{\sum_{m}T_{mn}\ast \zeta
_{m}^{-1}} \, \mathrm{e}^{-\sum_{m}S_{mn}\ast \xi _{m}^{-1}}.
\end{eqnarray}
Neglecting higher order correction terms we finally arrive at the
dressed energy equations (\ref{DEC}).
The pressure can readily be written in terms of the polylogarithm
function using (\ref{DEC}) and (\ref{p_all}).

\clearpage

\begin{figure}[t]
{{\includegraphics [width=0.80\linewidth]{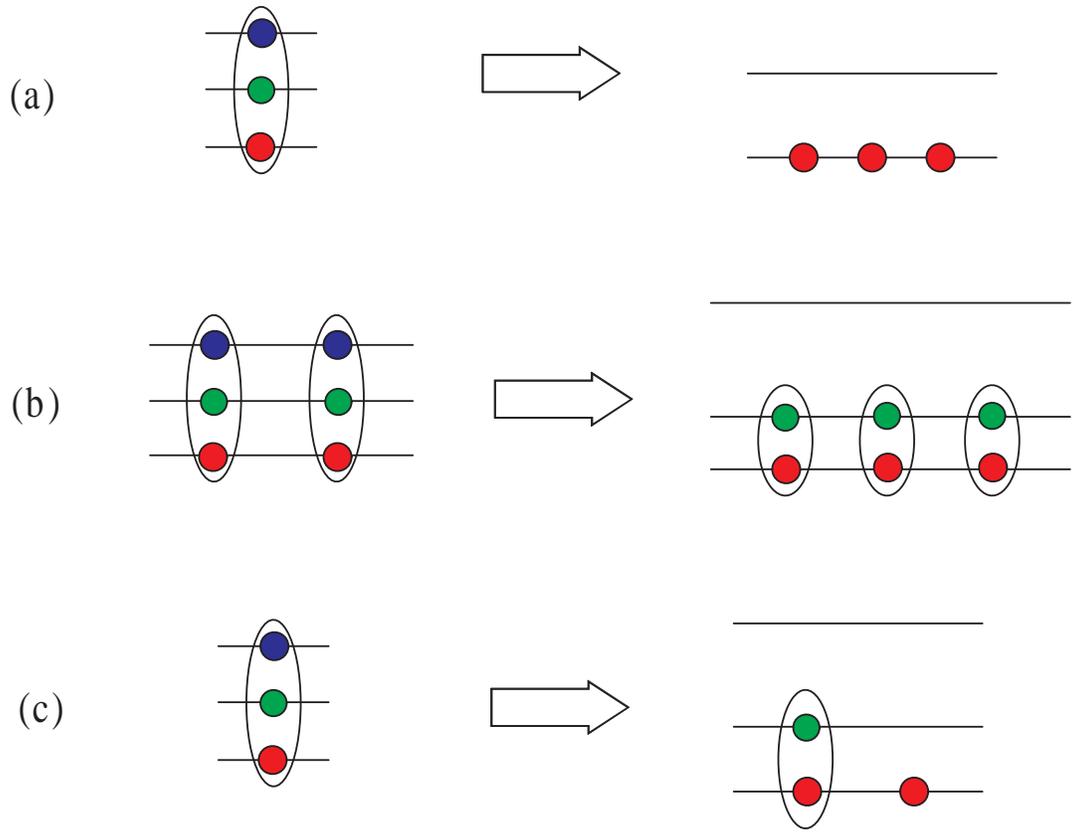}}}
\caption{(Color online) Phase transitions from states of trions into
(a) normal Fermi liquid, (b) fully-paired states and (c) the mixture
of pairs and unpaired fermions. The transitions are induced by
linear (a) and nonlinear ((b) and (c)) Zeeman splitting. Ellipses
denote charge bound states.} \label{fig:phases}
\end{figure}

\begin{figure}[t]
{{\includegraphics [width=1.0\linewidth]{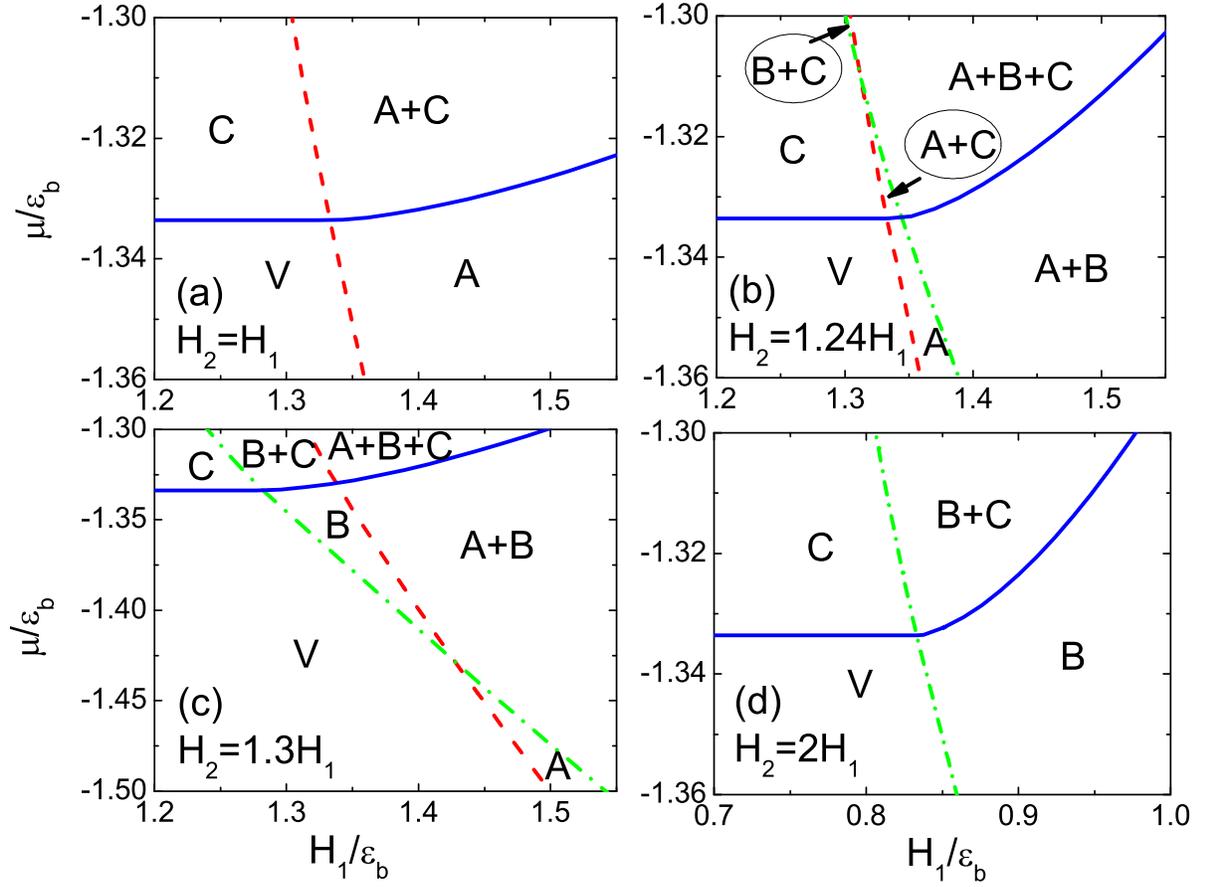}}} \caption{(Color
online) The $\mu -H$ phase diagrams at the temperature
  $T=0.001\varepsilon _{b}$ for (a) pure Zeeman splitting and
  unequally-spaced Zeeman splitting (b), (c) and (d).
  $V$ denotes the vacuum phase, $A$ denotes the unpaired fermion phase,
  $B$ denotes the paired phase and $C$ denotes the trion phase.
  The phase
  boundaries are determined by the equation of state (\ref{ppp}) which
  are consistent with the phase diagrams determined via the dressed energy
  equations, see the description in the text. }
\label{fig:mu-h}
\end{figure}

\begin{figure}[t]
{{\includegraphics [width=1.0\linewidth]{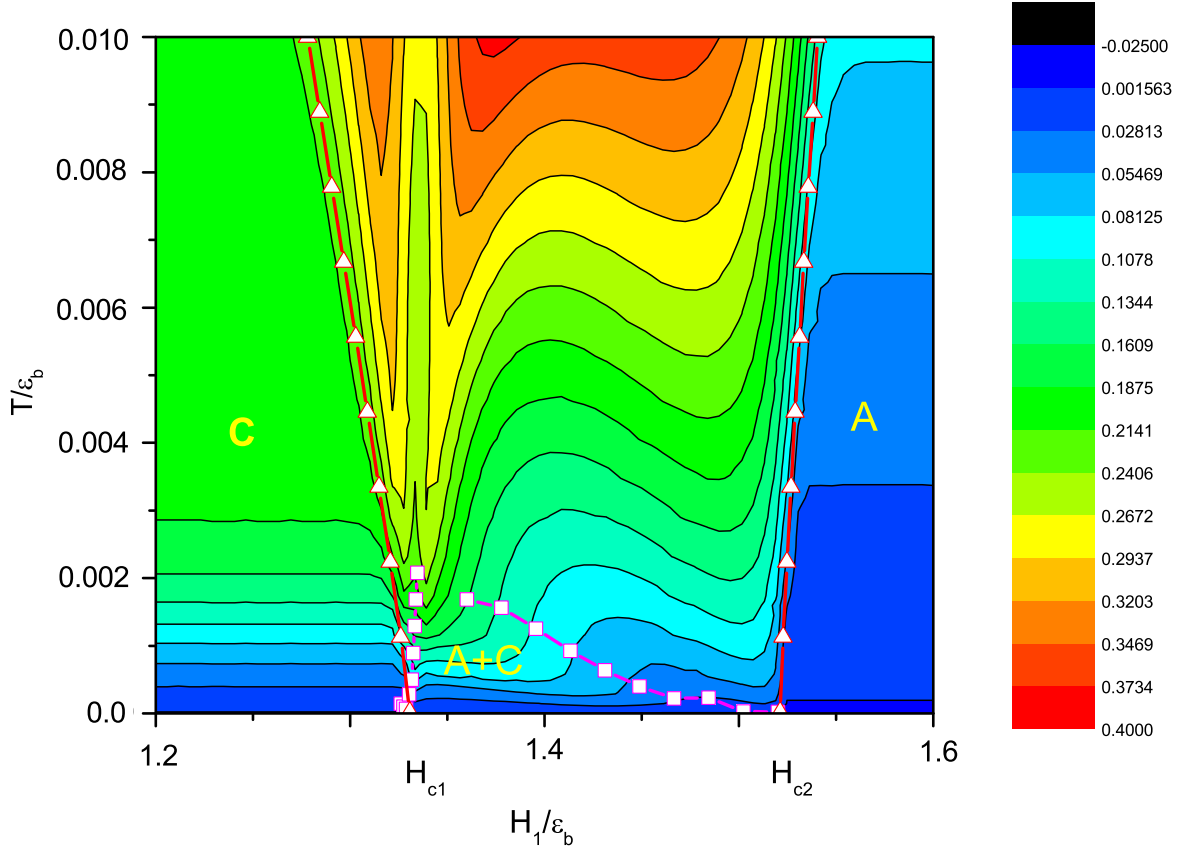}}} \caption{(Color
online) The specific heat $c_{v}$ in the $T$-$H_{1}$ plane for pure
Zeeman splitting with $H_{1}=H_{2}$ and total density $n$ fixed. An
asymmetric two-component TLL remains within a regime below the line
of squares between $ H_{c1}<H<H_{c2} $. The TLL of spin-neutral
trion states and the TLL of unpaired  fermionic atoms lie below the
left and right line of triangles, respectively. } \label{fig:Cv1}
\end{figure}

\begin{figure}[t]
{{\includegraphics [width=1.0\linewidth]{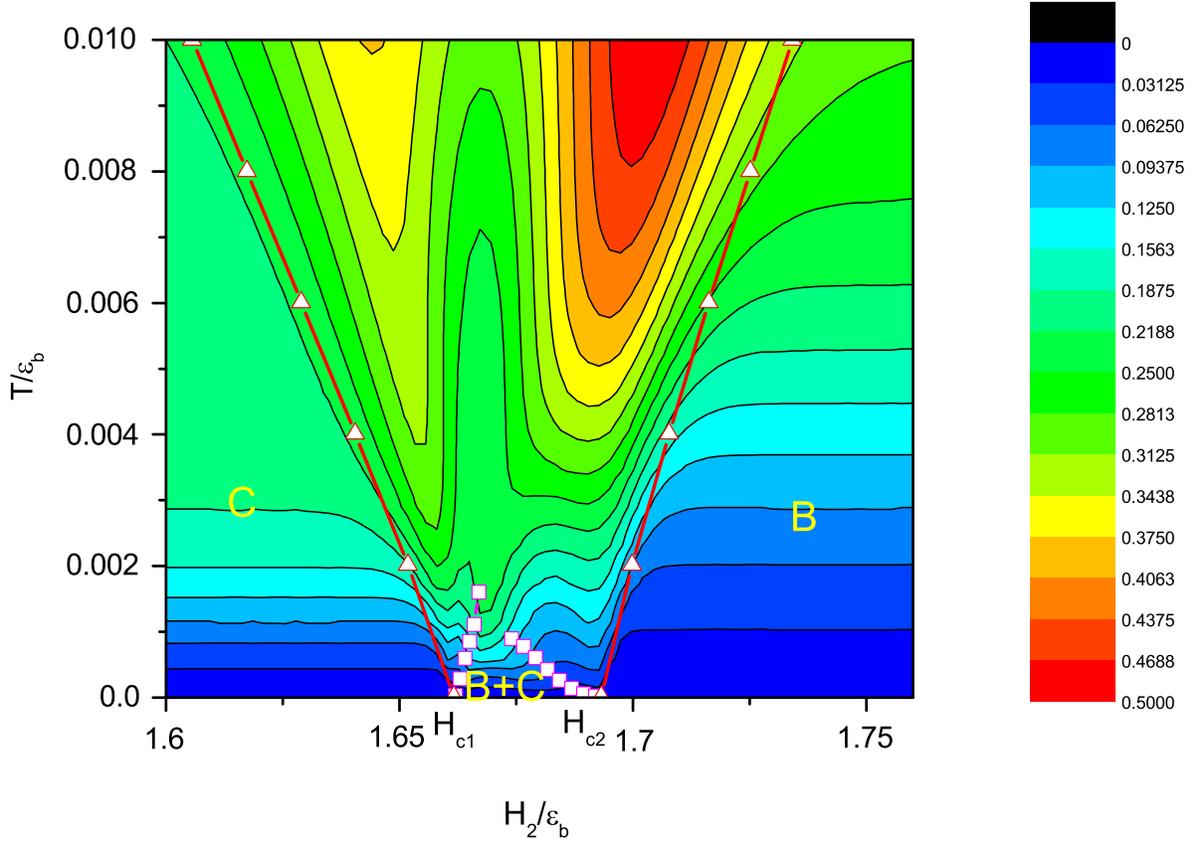}}} \caption{(Color
online) The specific heat $c_{v}$ in the $T$-$H_{2}$ plane for
$H_{2}=2H_{1}$ with total density $n$ fixed. An asymmetric
two-component TLL remains within a regime below the line of squares
between $H_{c1}<H<H_{c2}$. The TLL of spin-neutral trions and the
TLL of the composite pairs lie below the left and right lines of
triangles, respectively. } \label{fig:Cv2}
\end{figure}

\begin{figure}[t]
{{\includegraphics [width=1.0\linewidth]{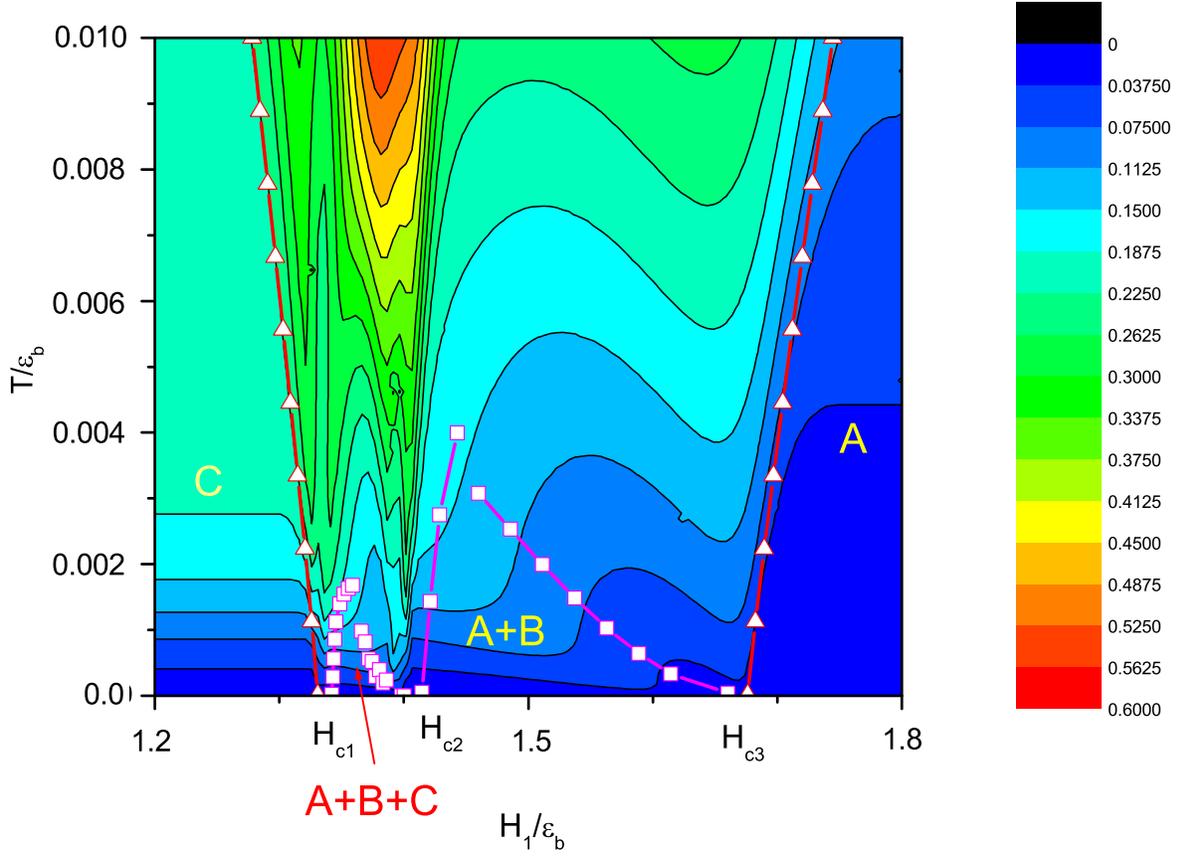}}} \caption{(Color
online) The specific heat $c_{v}$ in the $T$-$H_{1}$ plane for
$H_{2}=1.2H_{1}$ with total density $n$ fixed. An asymmetric
three-component TLL remains within a regime below the line of
squares between $ H_{c1}<H<H_{c2} $. An asymmetric two-component TLL
remains within a regime below the line of pink squares between
$H_{c2}<H<H_{c3}$. A TLL of spin-neutral trion states and a TLL of
unpaired  fermionic atoms lie below the left and right lines of
triangles, respectively.} \label{fig:Cv3}
\end{figure}

\begin{figure}[t]
{{\includegraphics [width=1.0\linewidth]{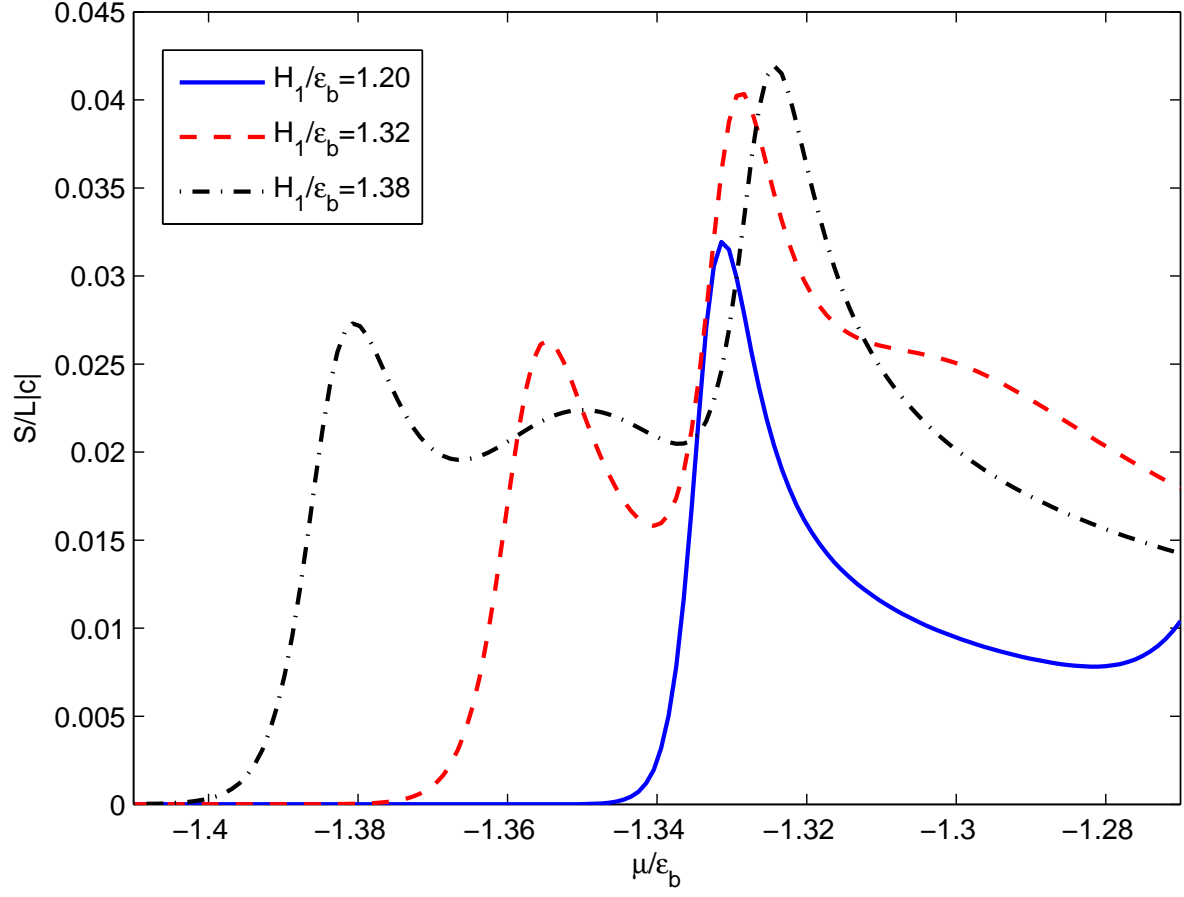}}} \caption{(Color
online) Entropy per unit length vs chemical potential with
$T/\varepsilon_b=0.005$ and $H_2=1.3 H_1$ for different
$H_1/\varepsilon_b$.  The entropy exhibits peaks in the phases of higher
density of states when the chemical potential passes the critical
points, see text. } \label{fig:S}
\end{figure}

\begin{figure}[t]
{{\includegraphics [width=1.0\linewidth]{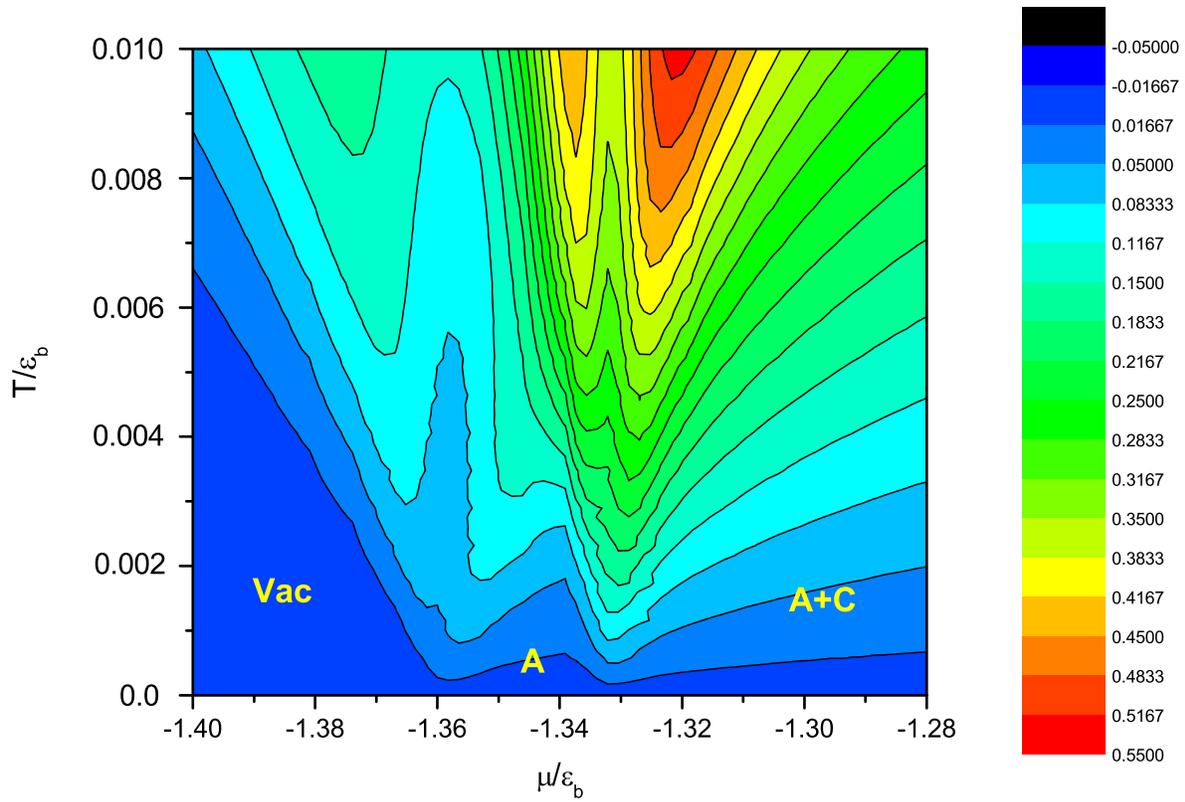}}} \caption{(Color
online) Specific heat phase diagram in the $T-\mu$ plane for
equally-spaced Zeeman splitting with $H_1=1.36 \varepsilon_{b}$. The
figure shows how the phase diagram extends out
from the zero temperature phase diagram. } \label{fig:c_v-mu}
\end{figure}

\end{widetext}

\end{document}